\documentclass[
aps,nofootinbib,
showpacs,showkeys,preprint
tightenlines,preprintnumbers,] {revtex4}

\usepackage{epsf,epsfig,subfigure,graphicx,amsmath,amssymb 
}
\usepackage{color}
\newcommand{\dis}[1]{\begin{equation}\begin{split}#1\end{split}\end{equation}}
\newcommand{\ie}{{\it i.e.~}}
\newcommand{\etal}{{\it et al.\,}}

\newcommand{\tev}{\,\textrm{TeV}}
\newcommand{\gev}{\,\textrm{GeV}}

\newcommand{\Mp}{M_{\rm P}}

\newcommand{\Uanom}{U(1)$_{\rm anom}$}

\newcommand{\UoR}{U(1)$_{\rm R}$}
\newcommand{\UEE}{U(1)$_{\rm EE}$}

\newcommand{\UKK}{U(1)$_{\rm KK}$}

\newcommand{\UoReq}{\rm U(1)_R}
 \newcommand{\UEEeq}{\rm U(1)_{EE}}
\newcommand{\UKKeq}{\rm U(1)_{KK}}

\def\sw0{{$\sin^2\theta_W^0$}}

\newcommand{\Z}{{\bf Z}}

\def\smg{{SU(3)$_C\times$SU(2)$_W\times$U(1)$_Y$}}
\def\E6{{\rm E_6}}

\def\EE8{{\rm E_8\times E_8'}}

\def\flip{SU$(5)_{\rm flip}$}

\def\one{{\bf 1}}

\def\five{{\bf 5}}
\def\ten{{\bf 10}}
\def\tenb{{\overline{\bf 10}}}

\def\fiveb{{\overline{\bf 5}}}

\begin{document}

\draft

\title{\Large\bf R-parity from string compactification}

\author{ Jihn E.  Kim}
\address
{Department of Physics, Kyung Hee University, 26 Gyungheedaero, Dongdaemun-Gu, Seoul 02447, Republic of Korea, and\\
Center for Axion and Precision Physics Research (Institute of Basic Science), KAIST Munji Campus, 193 Munjiro,  Daejeon 34051, Republic of Korea, and \\
  Department of Physics and Astronomy, Seoul National University, 1 Gwanakro, Gwanak-Gu, Seoul 08826, Republic of Korea 
}
 
\begin{abstract} 
In this paper, we embed the  $\Z_{4R}$ parity as a discrete subgroup of a global symmetry \UoR\,obtained from $\Z_{12-I}$ compactification of heterotic string $\EE8$.  A part of \UoR\,transformation is the shift of the anticommuting variable $\vartheta$ to $e^{i\alpha}\vartheta$ which necessarily incoorporates the transformation of internal space coordinate. Out of six internal spaces, we identify three U(1)'s whose charges are  denoted as $Q_{18},Q_{20}$, and $Q_{22}$. The \UoR~is defined as \UEE$\times$\UKK~where \UEE~is the part from $\EE8$ and \UKK~is the part generated by $Q_{18},Q_{20}$, and $Q_{22}$. We propose a method to define a \UoR~direction. The needed vacuum expectation values for breaking gauge U(1)'s except U(1)$_Y$ of the standard model carry \UoR~charge 4 modulo 4 such that \UoR~is broken down to $\Z_{4R}$ at the grand unification scale.   $\Z_{4R}$ is broken to $\Z_{2R}$ between the intermediate ($\sim 10^{11\,}\gev$) and the electroweak scales ($100\,\gev\sim 1\,\tev$).  The conditions we impose are proton longevity, a large top quark mass,  and acceptable magnitudes for the $\mu$ term  and neutrino masses.   
 
\keywords{R parity, U(1)$_{\rm R}$ symmetry,  $\Z_{4R}$,  String compactification, $\Z_{12-I}$ orbifold}
\end{abstract}
\pacs{11.25.Mj, 11.30.Er, 11.25.Wx, 12.60.Jv}
\maketitle


\section{Introduction}\label{sec:Introduction}
In supersymmetric (SUSY) extensions of the standard model (SM) and grand unified theories (GUTs), the proton longevity invites additional symmetries. The mostly discussed one    is the R-parity \cite{WeinbergPdecay,Sakai82}.\footnote{For a systematic study of matter parity in addition to R-parity, see Ref.  \cite{IbanezMatterP}  for example.}

``How is the current allocation of flavors realized?'' is the most urgent and also interesting one in the theoretical problems of the standard model (SM) of particle physics. Advocates of string theory for the heterotic string argue that string compactification is the most complete answer to this problem \cite{Candelas, Dixon2, Ibanez1,Tye87,Bachas87,Gepner87}.

String compactifications aim at obtaining (i) large 3D space, (ii) standard-like models with three families, and (iii) no exotics at low energy (or vectorlike representations if they exist). Regarding a solution to item (i), the string landscape scenario  is suggested \cite{LandscapeSuss}, predicting about $10^{500}$ vacua for a reasonable cosmological constant  (CC).  Regarding   item (ii),   the standard-like models from heterotic string  has been suggested from early days \cite{IKNQ,Munoz88} until recently \cite{Lykken96, PokorskiW99, Cleaver99,Cleaver01, CleaverNPB,Donagi02, Raby05, He05,Donagi05,He06, Donagi06, Blumenhagen06,Cvetic06, KimJH07,Faraggi07, Blumenhagen07, Cleaver07, Munoz07,Nilles08}. Model constructions are discussed in detail in \cite{LNP696,IbanezBk,RabyBk}.
  It has been suggested that by exploring the entire string landscape one might obtain statistical data which could lead to probabilistic experimental statements \cite{Schellekens04,Anastasopoulos}. Yet the clearest statement to date is that standard-like models are exceedingly rare \cite{Lust05,LandDouglas}. In addition, the flavor problem asks for a detail model producing the observed Cabibbo-Kobayashi-Maskawa (CKM) \cite{Cabibbo63,KM73} and Pontecorvo-Maki-Nakagawa-Sakada (PMNS) \cite{PMNS1,PMNS2} matrices. In the future, a more refined  statistical search, satisfying all the observed SM data, can be performed with the help of artificial intelligence (AI) program. At present, an AI program is not available for this purpose and hence we study this flavor problem analytically in the simplest orbifold compactification based on $\Z_{12-I}$.\footnote{Among the nine orbifolds of \cite{Dixon2}, we consider $\Z_{12-I}$ is the simplest one in the sense that it has only three fixed points.} Since the number of fields are over hundred in these standard-like models,   we simplify further by choosing GUT models to ease the analytical study.   Therefore, we require the following in addition to the above three items,
    \begin{itemize}
 \item[(iv)] Supersymmetry  is imposed at the grand unification (GUT) scale.
  
 \item[(v)] We consider GUT scale gauge groups as simple groups \cite{GG74,SO10GUT1,SO10GUT2,RamondE6} or semi-simple groups \cite{PS73,Barr82,DKN84}. 
   \end{itemize}
   
SUSY models have been widely used to introduce a mechanism for generating a hierarchically small electroweak (EW) scale compared to the GUT scale. Somewhere above the EW scale, therefore, SUSY must be broken since no superpartner has been observed up to a TeV scale \cite{LHCichep}. In the model,   SUSY breaking mechanism must be present. The well-known mechanism for SUSY breaking applicable to string compactification is the gaugino condensation \cite{Nilles82,DIN84,DRSW84}. Working in the SUSY breaking models from compactification, we require the gauge group at the GUT scale as $G_{\rm GUT}\times G_{\rm cond}$. Most probable $ G_{\rm cond}$ is SU(4)$'$ which can trigger SUSY breaking via gaugino condensation \cite{NillesSU4}. 

In SUSY models, R-parity $P_{\mathrm {R} }=(-1)^{3(B-L)+2S}$  dictates proton stability, where $B$ is baryon number,  $L$ is lepton number, and $S$ is spin. For a conserved R-parity, it is usually assigned to a subgroup of $B-L$. From string compactification, R-parity was calculated before in this framework \cite{KimKyae07R,KimJH07,Kappl09}. Because of dangerous dimension-5 operators, leading to proton decay, $\Z_{4R}$ has been proposed in contrast to $\Z_{2R}$  \cite{Maru01,Babu03,Lee11,Tamvakis12,KimZnR}. In this paper, we will present a detail study toward R-parity from the continuous symmetry U(1)$_R$. We will see that \UoR~is also constraining some couplings and hence helps to forbid some unwanted $\Delta B\ne 0$ operators.

GUTs from string compactification favor the flipped SU(5) semi-simple GUTs  \cite{Ellis89,KimKyae07,Huh09} and anti-SU(7) \cite{KimSU7}. For the simple group GUTs, SU(5), SO(10), and E$_6$, we need an adjoint representation to break the GUT groups down to the SM gauge group and it is impossible to obtain adjoint representation at the level 1 \cite{LNP696}. [Note, however, an adjoint representation of SO(10) was obtained in Ref. \cite{Tye96} at the level 3.] So, for simple studies at the level 1, anti-SU($N$) GUTs are relevant for phenomenological studies.\footnote{In Ref.  \cite{KimSU7}, anti-SU($N$) GUTs are defined as those that the GUT breaking is achieved by the anti-symmetric representations. In this definition, the flipped SU(5) is `anti-SU(5)'.}

In Sec. \ref{sec:Rparity}, after recapitulating the need for R-parities toward proton longevity, we discuss possibilities of embedding R parities in the global symmetry group \UoR~from string compactification. The specific example is presented in the flipped SU(5) model of Ref. \cite{Huh09}. Here the details of U(1) quantum numbers are presented for all the spectra. In Sec.   \ref{sec:UKK}, we define \UoR~global symmetry, including the R symmetry transformation of the anti-commuting variable $\vartheta$. The U(1) charges $Q_{18}, Q_{20}, $ and $Q_{22}$ are defined from three tori of the six compactified space. In Sec.  \ref{sec:NSinglets}, the neutral singlets which can obtain GUT scale vacuum expectation values (VEVs) are discussed. 
In Sec.   \ref{sec:Yukawa}, we discuss the resulting phenomenology. In Sec. \ref{sec:Vacuum}, we discuss the vacuum structure, leading to the above VEVs, and Sec. \ref{sec:Conclusion} is a conclusion.  In Appendix \ref{appA}, we present some details for obtaining massless fields and their  $Q_{18}, Q_{20}, $ and $Q_{22}$ quantum numbers.

\section{R-Parities}\label{sec:Rparity}

\begin{figure}[!t]
\includegraphics[width=0.25\textwidth]{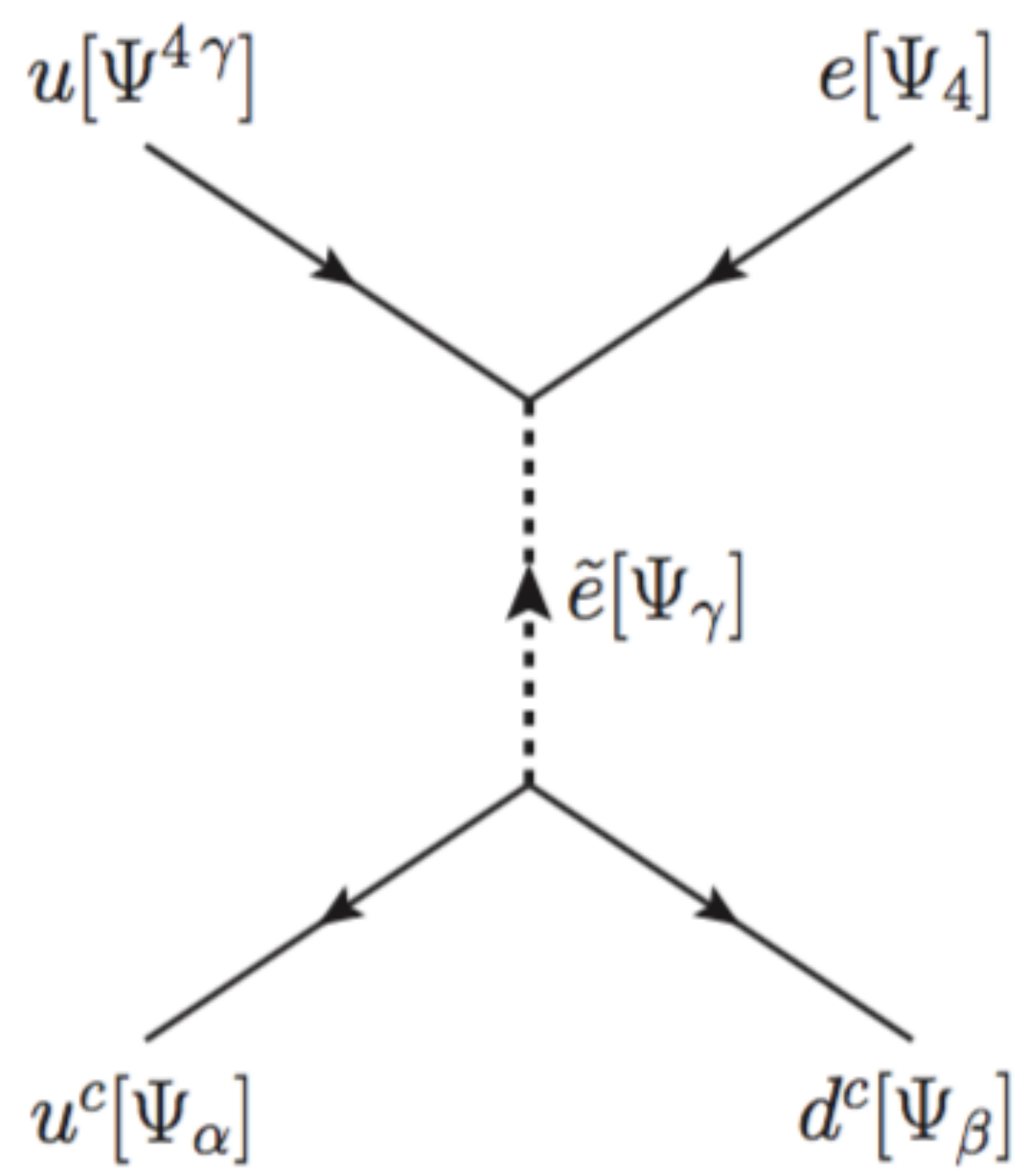} 
   \caption{A diagram for $\Delta B\ne 0$ without R parity. The cubic couplings in this diagram break R-parity.} \label{fig:Pdecay} 
\end{figure}

Beyond the SM (BSM), baryon (B) and lepton (L) numbers are broken. The degree of breaking depends on a BSM theory. The most widely discussed one that is also  relevant in our paper is the B violation in SUSY extensions of the SM. Supersymmetric standard models (SSMs) can start with the gauge symmetry \smg~with  B and L conserving dimension-4 operators, which has led to the R-party conservation and predicted the lightest supersymmetric particle as a dark matter candidate.  In standard-like models from string,  a vacuum  with R-parity was explicitly shown to exist first in \cite{KimJH07}.   The standard R-parity or $\Z_{2R}$ however have been known to be dangerous for the proton longevity due to the  dimension-5 operators \cite{WeinbergPdecay,Sakai82}.  Without R parity, of a dangerous dimension-5 operator  appears as shown in Fig. \ref{fig:Pdecay} \cite{Dermisek00}.

\begin{figure}[!t]
\includegraphics[width=0.45\textwidth]{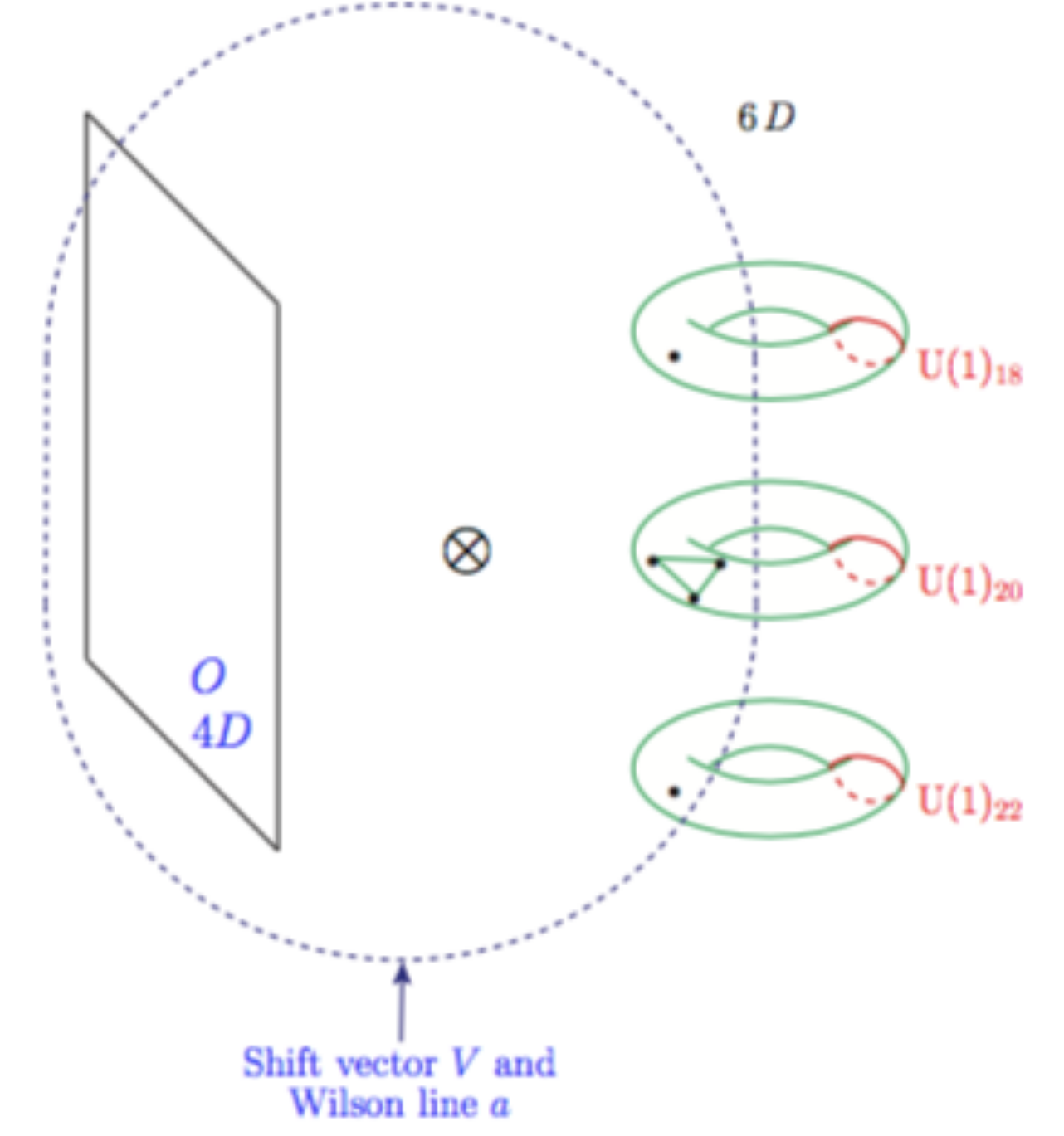} 
   \caption{Matching U(1)$_R$ with the rotation of the anticommuting variable $\vartheta$.  Four dimensional gauge groups are contained in the brane shown as the parallelogram, and three tori depict three \UKK's denoted as U(1)$_{18}$, U(1)$_{20}$, and U(1)$_{22}$. } \label{fig:U1R} 
\end{figure}

Without R-parity, forbidding dimension-5 B vioalting operators involves considering  all \flip~singlets which can obtain GUT scale VEVs in principle \cite{KimFlavor18}. Therefore, it will be economic in the discussion if the model contains some kind of R-parity. Dimension-5 B violating operators and the $\mu$ term are required to be suppressed but dimension-5 L violating Weinberg operator needs to be allowed \cite{WeinbergDim5}, but the $\mu$ problem must be resolved  \cite{KimNilles84,GMmu}. Considering the anomaly coefficients in SUSY field theory,  Lee \etal showed that non-R symmetries cannot be used to suppress the $\mu$ term \cite{Lee11}. Since we attempt to derive an R symmetry from string compactification that leads to consistent anomaly free models, consideration of anomaly coefficients from Lee \etal's point of view is not necessary. Anyway, we adopt their conclusion on $\Z_{4R}$ that the needed R-parity is a subgroup an \UoR~symmetry. So, let our \UoR~be a linear combination of \UEE~and \UKK, where  \UEE~is a U(1) from the gauge group $\EE8$ \cite{Gross85} and \UKK~is a U(1) from the internal space.  In Fig. \ref{fig:U1R}, gauge groups shown in 4D contain \UEE~and three tori depict three \UKK's denoted as U(1)$_{18}$, U(1)$_{20}$, and U(1)$_{22}$. For $\Z_3$ orbifolds, it was commented that dimension-3 $\mu$ term is forbidden \cite{Casas93}, but the intermediate scale $M_I$ generates the EW scale as $\sim M_I^3/\Mp^2$.  It was known that the common scale for breaking the PQ symmetry and supergravity is needed  \cite{Kim84W}. Also, for a multiple appearance of Higgs pairs, the democratic mass matrix, by some kind of fine tuning, always guarantees at least one massless pair of Higgs doublets \cite{KimKyaeNam17}. So, we may consider the  cases of discrete groups $\Z_4, \Z_6, \Z_8$ and $\Z_{12}$ of Ref. \cite{Lee11}. Illustration with $\Z_{4R}$ from \UKK~of $\Z_{12-I}$ orbifold can be applicable to the other cases also. 

\begin{figure}[!t]
\includegraphics[width=0.35\textwidth]{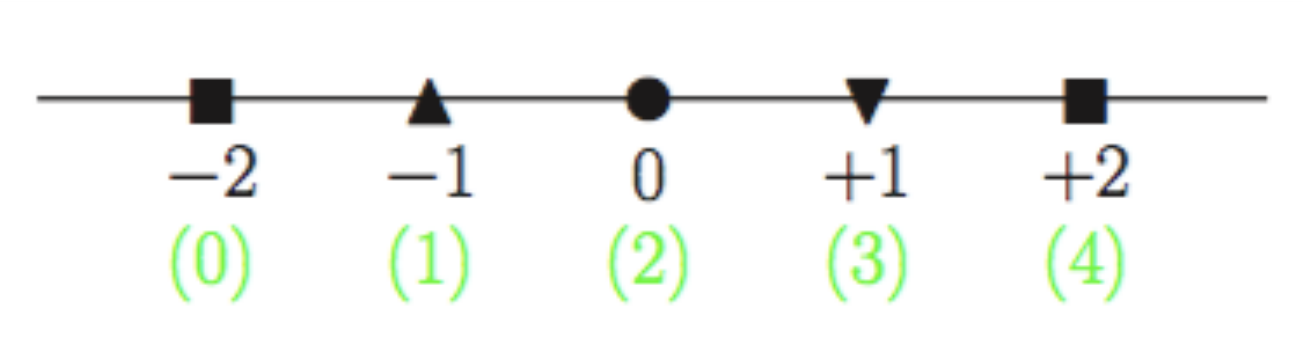} 
   \caption{$\Z_{4R}$ quantum numbers in the region $[-2,+2]$.  Numbers in the region $[0,4]$ are shown  in the brackets.} \label{fig:Rparity} 
\end{figure}

 Let us consider the following operators, relevant for the dimension-5 proton decay and neutrino mass operators,
\dis{
&W^{\Delta B}\equiv \tenb_m \tenb_m \tenb_m \five_m ,\\
&W^{\nu\,\rm mass}\equiv \five_m \five_m \fiveb_{H_u} \fiveb_{H_u}, 
}
where the subscripts $m$ and $H$ denote matter fields and Higgs fields, respectively. If an operator is present in the superpotential, \UKK~transformations of the fields of an operator is cancelled by the transformation of the anti-commuting variable $\vartheta$. Under certain normalization, the superpotential is required to have +2 units of the \UKK~charge. Since the rotation angle of variable $\vartheta$ can be taken as the negative of the previous transformation,  --2 units of the \UKK~charge must be allowed also as illustrated in Fig. \ref{fig:Rparity}. So, we have a $\Z_4$ symmetry $-2\equiv +2$, \ie minimally we require $\Z_{4R}$ symmetry when we consider the global transformation of  $\vartheta$.  The $\Z_{4R}$ quantum numbers can be labeled as those in green color, and the black number assignment is identical to those of green colors.  Under $\Z_{4R}$, the superpotential $W$ leading to proton decay operator and the $\mu$ term are required to carry $+4\equiv 0$ units which are then forbidden by U(1)$_R$, and the superpotential for  neutrino mass   operator carries +2 units which is allowed by U(1)$_R$. These can be satisfied with the matter charges +1 and the Higgs charges 0, for example. In string compactification, the realization may be more complex because one must take into account the sectors where these fields appear.

The R-parity is a discrete subgroup of \UoR,
\dis{
\UoReq\subset \UEEeq\otimes \UKKeq.
}
Superpotential carries +2 (modulo 4) units of \UoR. On the other hand, the integrand under $d^2\vartheta\,d^2\bar{\vartheta} $ carries +4 (modulo 4) units of \UoR. 

\subsection{Model}\label{subsec:Model}

The shift vector $V$ and Wilson line are
\dis{
&V=\left(0,0,0,0,0;\frac{-1}{6},\frac{-1}{6},\frac{-1}{6}\right)\left(0,0,0,0,0;\frac{1}{4},\frac{1}{4},\frac{-2}{4}\right)',\\[0.3em]
 &a=\left(\frac{2}{3},\frac{2}{3},\frac{2}{3},\frac{2}{3},\frac{2}{3};0,\frac{-2}{3},\frac{2}{3} \right)
 \left(\frac{2}{3},\frac{2}{3},\frac{2}{3},\frac{2}{3},0;\frac{-2}{3},0,0\right)' \label{eq:modelV}
}
which gives the 4D gauge group SU(5)$\times$SU(5)$'\times$SU(2)$'\times$U(1)$^7$. 

\subsection{U(1) charges of $\EE8$}\label{subsec:UEE}

The U(1)$_X$ charge of \flip~is
\dis{
X=(-2,-2,-2,-2,-2;0^3)(0^8)'
}
and $Q_{\rm anom}$ is given by
\dis{
Q_{\rm anom}= \frac{1}{126}(84Q_1+147Q_2-42Q_3-63Q_5-9Q_6),
}
where
\dis{
&Q_1= (0^5;12, 0, 0)(0^8)' ,\\
&Q_2= (0^5;0,12,  0)(0^8)'  ,\\
&Q_3= (0^5;0,0,12)(0^8)'  ,\\
&Q_4=(0^8)  (0^4,0;12, -12, 0)',\\
&Q_5=(0^8)  (0^4,0;-6,-6,12)',\\
&Q_6=(0^8)  (-6,-6,-6,-6,18;0,0,6)' .
}
Any combination of $Q_{i}$ for $i=1,2,\cdots,6$ can be used for \UEE.

\subsection{\UKK}\label{subsec:UKK}

In this paper, compactification of six internal dimensions (coordinate $y$) is specified as three two-tori. So, any effective field $\Phi$ can be a function of $\Phi(x,y)$. To an observer in the 4D $x$ space, gauge symmetries in $y$ is global symmetries. So, the \UoR~symmetry we discuss must be a gauge symmetry in $y$ variable in the three tori. Let the radii of three tori be (radius)$_1$, (radius)$_2$, and (radius)$_3$, respectively. Then, the six internal coordinates are parametrized by  (radius)$_1e^{-i\varphi_1}$,  (radius)$_2e^{-i\varphi_2}$, and  (radius)$_3e^{-i\varphi_3}$. The right mover coordinates are given by\footnote{$\pm$ are $\pm\frac12$.}
\dis{
(\oplus|+++), (\oplus|+--), (\oplus|-+-), (\oplus|--+), 
\label{eq:Lmover}
}
and
\dis{
(\ominus|-++), (\ominus|+-+), (\ominus|++-), (\ominus|---), 
\label{eq:Rmover}
}
where (\ref{eq:Lmover}) is called R-handed (with $\oplus$) and  (\ref{eq:Rmover}) is called L-handed (with $\ominus$) . Gauge transformtions in the $y$-space rotate $\varphi$ angles and the generator for this rotations are called $Q_{18}, Q_{20}$, and $Q_{22}$, respectively, specifying the ranks of the total local group in addition to 16 of $\EE8$. We normalize the charges as
\dis{
&Q_{18}={\rm diag.}\left(2,0,0\right),\\[0.3em]
&Q_{20}={\rm diag.}\left(0,2, 0\right),\\[0.3em]
&Q_{22}={\rm diag.}\left(0,0,2\right).\label{eq:UoneRs}
}

In the standard-like models in this compactification leading to \smg$\times U(1)^{n}$, we have $n=15$. In the flipped SU(5) of \cite{Huh09},  SU(5)$\times$SU(5)$'\times$SU(2)$'\times$U(1)$^n$, we have $n=10$. To break all U(1)'s in the standard-like models, we need 16 independent vacuum expectation values (VEVs) of the Higgs fields.

To obtain superpotentials of massless superfields, the couplings must have $+2$ units of \UoR~charge. An appropriate combination of three U(1)'s in Eq.  (\ref{eq:UoneRs}) can be used for \UKK.
 
\subsection{Multiplicity}\label{subsec:Multiplicity}
We are interested in the multiplicity of massless states, $M^2=M_L^2+M_R^2=0$ for $M_L^2=M_R^2=0$,
\dis{
M_L^2&=\frac{(P+kV_f)^2}{2}+  \tilde{c}_k ,\\[0.2em]
M_R^2&=\frac{(s+k\phi' )^2}{2}+ c_k ,
}
where $\phi'=(0;\phi), s=(\oplus\textrm{ or }\ominus;\tilde{s})$, and
  $2\tilde{c}_k$ and $2c_k$ are  listed  in Appendix \ref{appA}. 
With $P$'s in Ref. \cite{Huh09}, one can check that $M_L^2=0$ is satisfied. The $M_R^2=0$ condition is used to obtain the chirality.

For the $\Z_{12-I}$ model of (\ref{eq:modelV}), the multiplicity of the massless spectrum    in  the sector $T_k$ sectors are
\dis{
{\cal P}_k(f)=\frac{1}{12\cdot 3}\sum_{l=0}^{11} \tilde{\chi}(\theta^k,\theta^l) e^{i\,2\pi l\,\Theta_k}\label{eq:multiplicity}
}
where $f(=\{f_0,f_+,f_- \})$ denote twisted sectors associated with $k V_f=kV, k(V+a), k(V-a)$. The phase $\Theta_k$ is given by
\dis{
\Theta_k=\sum_{i} (N_i^L-N_i^R)\hat{\phi}_i +(P+kV_f)\cdot V_f-( {s}+k\phi)\cdot \phi -\frac{k}{2}(V^2-\phi^2) ,\label{PhaseMult}
}
where $\frac{1}{2}(V^2-\phi^2)=\frac{ 2}{24}$, and $\hat{\phi}_j =\phi_j$ and $\hat{\phi}_{\bar{j}} =-\phi_j$.  For $k=0,3,6,9$, ${\cal P}_k(f_0)={\cal P}_k(f_+)={\cal P}_k(f_-)$ and the overall coefficient in Eq. (\ref{eq:multiplicity}) is $\frac{1}{12}$ instead of $\frac{1}{36}$, and we require in addition,
\dis{
P\cdot a=0 \textrm{\rm ~mod } Z~ \textrm{\rm ~in the } U, T_3, T_6, T_9~{\rm sectors}.
}
Note that the four entry $s$ and the three entry $\tilde{s}$ with the relation $s=(\oplus\textrm{ or }\ominus; \tilde{s}) $ such that  $\oplus$ or $\ominus$ is chosen to make the total number of minus signs even.
For the subsector $f=0$, \ie for $T^k_0$, from the masslessness condition, $2\tilde{c}_k=\sum_{i} (N_i^L)\hat{\phi}_i +P\cdot V  +\frac{k}{2}V^2 $, we have
\dis{
\Theta_k&=\sum_{i} (N_i^L-N_i^R)\hat{\phi}_i +P\cdot V - {s} \cdot \phi +\frac{k}{2}(V^2-\phi^2) \\
&=\sum_{i} (N_i^L-N_i^R)\hat{\phi}_i +P\cdot V - {s} \cdot \phi +\frac{k}{12} .\label{ThetaModel}
}
 
In the $\Z_N$ orbifold,   the $k=\frac{N}{2}$ sector is self-conjugate in a sense. The reason is the following. In the prime orbifolds, there is no sector $T_{N/2}$.  In the nonprime orbifolds, there is always a sector $T_{N/2}$. We do not consider the sector $T_{N}$ in this notation. Instead, we consider the untwisted sector $U$.\footnote{The untwisted sector corresponds to $k=0$ in Eq. (\ref{eq:multiplicity}). For the untwisted sector  $U_i$ for the torus $ i(=1,2,3)$, it is a closed string moving in the bulk of torus  $U_i$.} Then, there are $N-1$ twisted sectors where we consider only $k\le N/2$. With $k<N/2$, effectively we encompass $2(N-1)$ twisted sectors. The remaining two twised sectors are in $T_{N/2}$.  

\subsection{Selection rule in $\Z_N$}\label{subsec:Selection}
One imporant selection rule of Yukawa couplings is to satisfy $\Z_N$ invariance both for the L- and R-sectors. This amounts to satisfying the sum of phases $\sum_i\Theta_i=0$ modulo 12 for dimension $n$ superpotential,  $W_n\propto \prod_{i=1}^n\Phi_i$.

\section{U(1) charges $Q_{18},Q_{20}$ and $Q_{22}$, and the SM fields}\label{sec:UKK}

In our $\Z_{12-I}$ model of Subsec. \ref{subsec:Model},   we use the following normalization,
\dis{
Q_{18}:(2, 0,0),~Q_{20}:(0,2, 0),~Q_{22}:(0,0,2).
}

\subsubsection{Untwisted sector}

The multiplicity of the massless spectrum  in  the untwisted sector $U_i$ occurs with $P\cdot V=\frac{n_i}{12}$ where $n_i=5,4,1$ for the tori index $i=1,2,3$.   The $U_3$ fields in Table I has $P\cdot V=\frac{1}{12}$. This must be canceled by the phase of right movers. Note that $\tilde{\chi}(\theta^0,\theta^l)  $ of Eq. (\ref{eq:multiplicity}) is 3, and 
\dis{
\Theta_{U_3}= \frac{1}{12}- \tilde{s}_i \cdot \phi 
}
The phase in the 3rd torus
$\Theta_{U_3}=0$ is achieved by $\tilde{s}=(\ominus;+-+)$ such that $ \tilde{s}_i \cdot \phi =\frac{1}{12}$ where $\phi=(\frac{5}{12},\frac{4}{12},\frac{1}{12} )$. 
It is L-handed, \ie $\ominus$, in our definition of the handedness, and obtain
\dis{
Q_{18,20,22}=+1,-1,+1,
}
respectively, which are listed in Table I.

\subsubsection{Twisted sectors}

 We will present most  twisted sector fields in detail in Appendix A except for the Higgs fields needed in the \flip:  $H_{u,d}$ and $\ten_{+1\,H}$ and $\tenb_{-1\,H}$.

\vskip 0.3cm
\noindent{\bf Sector $T_4^0$}: Here, two families ($\xi_{2,3}, \bar{\eta}_{2,3}, \mu^c,\tau^c$) of Table I appear, which are calculated in detail in Appendix \ref{appA}. 
   
   \vskip 0.3cm
\noindent{\bf Sector $T_6$}: We locate the light Higgs doublets in this sector.  From Eq. (\ref{eq:multiplicity}), multiplicities in the $T_6$ sector is calculated with the following $\tilde{\chi}(\theta^6,\theta^j)$,\footnote{Use $s$ in the mass relation and use $\tilde{s}$ in the phase calculation.}
\dis{
\tilde{\chi}(\theta^6,\theta^j)=\left\{\begin{array}{lrrrrrrrrrrrr} 
j=&0,&1,&2,&3,&4,&5,&6,&7,&8,&9,&10,&11 \\
&16,&1,&1,&4,&1,&1,&16,&1,&1,&4,&1,&1 \end{array}\right.\label{eq:MultT6}
}
In   $T^6$, we have  from (\ref{ThetaModel})
\dis{
\Theta_6& =\sum_{i} (N^L_i-N_i^R)\hat{\phi}_i - \tilde{s}  \cdot\phi+P\cdot V+\frac12.\label{Theta6}
}
For $H_{uL}=(\underline{+1\,0\,0\,0\,0};\,-1\,0\,0)(0^5;-1\,0\,+1)'$, we have $P\cdot V=\frac{+5}{12}$. Since $6\phi=(\frac12,0,\frac12)$, the masslessness condition is satisfied for $s=( \oplus\textrm{ or }\ominus;-,\pm,-)$, and we obtain the following multiplicity  
\dis{
 \begin{array}{ccccc} 
 {s}  & N_i\hat{\phi}_i,  & \tilde{s}  \cdot\phi,&\Theta_6, &{\rm Multiplicity}\\[0.3em]
 (\oplus|-+-):&0, &\frac{-1}{12}, & 0,&4\cdot {H_{uR}}\\[0.3em] 
 (\ominus|---):&0, &\frac{-5}{12}, &\frac{+4}{12},& 2\cdot H_{uL}
 \\[0.3em] 
  \end{array}  \label{eq:HiggsT6}
}
   Similarly, we obtain $H_{d}$'s, and there result the following Higgs doublets from $T_6$,
 \dis{
{\color{red} 2} \cdot  H_{uL}(-1,-1,-1)+{\color{red} 2}\cdot  H_{dL}(-1,-1,-1)+4\cdot H_{uR}(-1,+1,-1)+ 4\cdot H_{dR}(-1,+1,-1)  .\label{eq:AllHs}
 }
Since the R-handed fields of (\ref{eq:AllHs}) do not contribute to the superpotential for Yukawa couplings,   we list only the L-handed fields, colored red in Eq. (\ref{eq:HiggsT6}), whose 
 $Q_{18,20,22}$ quantum numbers  are listed in Table I. 
  
\begin{table}[t!]
\begin{center}
\begin{tabular}{@{}lc|cc|c|cccccc|c|ccc@{}} \toprule
 &State($P+kV_0$)&$~\Theta_i~$ &${\bf R}_X$(Sect.)&$Q_R$ &$Q_1$&$Q_2$ &$Q_3$ &$Q_4$ &$Q_5$ &$Q_6$ & $Q_{\rm anom}$& $Q_{18}$& $Q_{20}$& $Q_{22}$ \\[0.1em] \colrule
 $\xi_3$  & $(\underline{+++--};--+)(0^8)'$&$0$ &$\tenb_{-1}(U_3)$&$-5$ &$-6$ & $-6$ & $+6$ & $0$ & $0$ & $0$ & $-13$ & $+1$ & $-1$& $+1$ \\
$\bar{\eta}_3$  & $(\underline{+----};+--)(0^8)'$&$0$ &$\five_{+3}(U_3)$&$-5$  & $+6$ & $-6$ & $-6$ & $0$ & $0$ & $0$ & $-1$ & $+1$& $-1$& $+1$ \\
$\tau^c$  & $({+++++};-+-)(0^8)'$& $0$ &$\one_{-5}(U_3)$ &$-5$& $-6$ & $+6$ & $-6$ & $0$ & $0$ & $0$ & $+5$ & $+1$& $-1$ & $+1$  \\
$\xi_2$  & $(\underline{+++--};-\frac{1}{6},-\frac{1}{6},-\frac{1}{6})(0^8)'$& $\frac{+1}{4}$ &$\tenb_{-1}(T_4^0)$& $-5$ &$-2$ & $-2$ & $-2$ & $0$ & $0$ & $0$ & $-3$ & $-1$& $-1$ & $-1$\\
$\bar{\eta}_2$  & $(\underline{+----};-\frac{1}{6},-\frac{1}{6},-\frac{1}{6})(0^8)'$&$\frac{+1}{4}$ &$\five_{+3}(T_4^0)$& $-5$   & $-2$ & $-2$ & $-2$ &$0$ & $0$ & $0$ & $-3$ & $-1$& $-1$ & $-1$\\
$\mu^c$  & $({+++++};-\frac{1}{6},-\frac{1}{6},-\frac{1}{6})(0^8)'$&$\frac{+1}{4}$ & $\one_{-5}(T_4^0)$& $-5$  & $-2$ & $-2$ & $-2$ &$0$ & $0$ & $0$ & $-3$ & $-1$& $-1$ & $-1$\\
$\xi_1$  & $(\underline{+++--};-\frac{1}{6},-\frac{1}{6},-\frac{1}{6})(0^8)'$&$\frac{+1}{4}$ &$\tenb_{-1}(T_4^0)$& $-5$  &$-2$ & $-2$ & $-2$ & $0$ & $0$ & $0$ & $-3$ &  $-1$& $-1$ & $-1$\\
$\bar{\eta}_1$  & $(\underline{+----};-\frac{1}{6},-\frac{1}{6},-\frac{1}{6})(0^8)'$&$\frac{+1}{4}$ &$\five_{+3}(T_4^0)$& $-5$   &$-2$ & $-2$ & $-2$ & $0$ & $0$ & $0$ & $-3$ & $-1$& $-1$ & $-1$\\
$e^c$  & $({+++++};-\frac{1}{6},-\frac{1}{6},-\frac{1}{6})(0^8)'$&$\frac{+1}{4}$ & $\one_{-5}(T_4^0)$& $-5$  &$-2$ & $-2$ & $-2$ &$0$ & $0$ & $0$ & $-3$ & $-1$& $-1$ & $-1$\\[0.2em]
\hline
 $H_{uL}$  & $(\underline{+1\,0\,0\,0\,0};\,0\,0\,0)(0^5;\frac{-1}{2}\,\frac{+1}{2}\,0)'$&$\frac{+1}{3}$ &$2\cdot \five_{-2}(T_6)$& $-2$  & $0$ & $0$ & $0$ & $-12$ & $0$ & $0$ & $0$ & $-1$& $-1$& $-1$   \\
 $H_{dL}$  & $(\underline{-1\,0\,0\,0\,0};\,0\,0\,0)(0^5;\frac{+1}{2}\,\frac{-1}{2}\,0 )'$&$\frac{+1}{3}$ &$ 2\cdot \fiveb_{+2}(T_6)$&$-2$ &$0$ & $0$ & $0$ & $+12$ & $0$ & $0$ & $0$ & $-1$ & $-1$& $-1$   \\ [0.2em]
 \botrule
 \end{tabular} \label{tab:SMqn}
\end{center}
\caption{U(1) charges of matter fields in the SM.  $\xi_i$ and $\bar{\eta}_i$ contain  the left-handed quark and lepton doublets, respectively, in the i-th family.}
\end{table}
\section{BSM fields: Neutral singlets}\label{sec:NSinglets} 

The BSM fields  must be neutral singlets and vector-like representations under \smg.
The state vectors containing neutral singlets are presented in the second column of Table II. All these neutral singlets appear in the twisted sectors. Neutral singlets are divided into two classes: one contained in the \flip~non-singlets $\Sigma^*=\tenb_{-1}$ and $\Sigma=\ten_{+1}$, and the other \flip~singlets $\sigma$'s. The VEVs of neutral components in $\Sigma^*$ and $\Sigma$ are needed to break the \flip\,down to the SM gauge group. In this section, we present the details on  $\Sigma^*$ and $\Sigma$.  \flip~singlets will be discussed in Appendix A.

\begin{table}[t!]
\begin{center}
\begin{tabular}{@{}lc|ccc|c|cccccc|c|ccc@{}} \toprule
 &State($P+kV_0$) &$\Theta_i$ & $(N^L)_j$&${\cal P}\cdot {\bf R}_X$(Sect.)&$Q_R$ &$Q_1$&$Q_2$ &$Q_3$ &$Q_4$ &$Q_5$ &$Q_6$ & $Q_{\rm anom}$& $Q_{18}$& $Q_{20}$& $Q_{22}$  \\[0.1em] \colrule
 $\Sigma^*_1$  & $(\underline{+++--};0^3)(0^5;\frac{-1}{4}\,\frac{-1}{4}\,\frac{+2}{4})'$ &$0$ &$2(1_1) $ & $2\,\tenb_{-1}(T_3)_L$ &$+4$& $0$ & $0$ & $0$ & $0$ & $+9$ & $+3$ & $\frac{-33}{7}$ & $-1 $& $+1 $ & $-1 $  \\
$\Sigma^*_1$  & $(\underline{+++--};0^3)(0^5;\frac{-1}{4}\,\frac{-1}{4}\,\frac{+2}{4})'$ &$\frac{+2}{3}$ &$1(1_{3})$ & $1\,\tenb_{-1}(T_3)_L$ &$+4$& $0$ & $0$ & $0$ & $0$ & $+9$ & $+3$ & $\frac{-33}{7}$ & $-1 $& $+1 $ & $-1 $  \\
${\Sigma}_2$  & $(\underline{++---};0^3)(0^5;\frac{+1}{4}\,\frac{+1}{4}\,\frac{-2}{4})'$& $0$ &$2(1_{\bar{1}})$&$2\, \ten_{+1}(T_3)_L$ &$-4$&$0$ & $0$ & $0$ & $0$ & $-9$ & $-3$ & $\frac{+33}{7}$ & $-1  $& $-1 $ & $-1 $ \\
 ${\Sigma}_2$  & $(\underline{++---};0^3)(0^5;\frac{+1}{4}\,\frac{+1}{4}\,\frac{-2}{4})'$& $\frac{+1}{3}$ &$1(1_{\bar{3}})$&$1\, \ten_{+1}(T_3)_L$ &$-4$&$0$ & $0$ & $0$ & $0$ & $-9$ & $-3$ & $\frac{+33}{7}$ & $-1  $& $-1 $ & $-1 $ 
 \\[0.2em]
\hline
$\sigma_1$  & $(0^5;\frac{-2}{3}\, \frac{-2}{3}\, \frac{-2}{3})(0^8)'$&$\frac{+1}{4}$ & $0$&$2\cdot\one_{0}(T_4^0)$  &$-14$& $-8$ & $-8$ & $-8$ & $0$ & $0$ & $0$ & $-12$ & $-1  $& $-1 $ & $-1 $  \\
$\sigma_2$  & $(0^5;\frac{-2}{3}\, \frac{+1}{3}\, \frac{+1}{3})(0^8)'$&$0$ &$3(1_{\bar{1}})$& $3\cdot \one_{0}(T_4^0)$ &$-2$& $-8$ & $+4$ & $+4$ & $0$ & $0$ & $0$ & $-2$ &  $-1  $& $-1 $ & $-1 $  \\
$\sigma_3$  & $(0^5;\frac{1}{3}\, \frac{-2}{3}\, \frac{1}{3})(0^8)'$&$0$ &$3(1_{\bar{1}})$&$3\cdot\one_{0}(T_4^0)$ &$-2$&$+4 $ & $-8$ & $+4$ & $0$ & $0$ & $0$ & $-8$ &    $-1  $& $-1 $ & $-1 $  \\
$\sigma_4$  & $(0^5;\frac{1}{3}\, \frac{1}{3}\, \frac{-2}{3})(0^8)'$&$0$ &$3(1_{\bar{1}})$&$3\cdot\one_{0}(T_4^0)$  &$-2$& $+4$ & $+4$ & $-8$ & $0$ & $0$ & $0$ & $+10$ &   $-1  $& $-1 $ & $-1 $  \\
$\sigma_5$  & $(0^5;0\,1\,0)(0^5;\frac{1}{2}\,\frac{-1}{2}\,0)'$&$\frac{+1}{2}$ &$0$& $2\cdot\one_{0}(T_6) $ &$+4$& $0$ & $+12$ & $0$ & $+12$ & $0$ & $0$ & $+14$ &   $-1  $& $-1 $ & $-1 $  \\
$\sigma_6$  & $(0^5;0\,0\,1)(0^5;\frac{-1}{2}\,\frac{1}{2}\,0)'$&$\frac{+1}{2}$ &$0$&$2\cdot\one_{0}(T_6) $ &$+4$&$0 $ & $0$& $+12$ & $-12$ & $0$ & $0$ & $-4$ &  $-1  $& $-1 $ & $-1 $  \\
$\sigma_7$  & $(0^5;0\,-1\,0)(0^5;\frac{-1}{2}\,\frac{1}{2}\,0)'$&$\frac{+1}{2}$ &$0$&$2\cdot\one_{0}(T_6)_R$  &$+8$& $0$ & $+12$ & $0$ & $+12$ & $0$ & $0$ & $+14$ &   $-1  $& $+1 $ & $-1 $  \\
$\sigma_8$  & $(0^5;0\,0\,-1)(0^5;\frac{1}{2}\,\frac{-1}{2}\,0)'$&$\frac{+1}{2}$ &$0$& $2\cdot \one_{0}(T_6)_R$ & $+8$ & $0$&$0$ & $+12$ & $-12$ & $0$ & $0$ & $-4$ &   $-1  $& $+1 $ & $-1 $  \\
$\sigma_{11}$ & $(0^5;\frac{-1}{2}\,\frac{-1}{2}\,\frac{-1}{2})(0^5;\frac{3}{4}\,\frac{-1}{4}\,\frac{-1}{2})'$& $\frac{+2}{3}$ & $2(1_1+1_3,1_{\bar{1}}+ 1_{\bar{3}} )$& $ 2\cdot\one_{0}(T_3)$ &$-9$& $-6$ & $-6$ & $-6$ & $+12$ & $-9$ & $-3$ & $\frac{-30}{7}$ &  $+1  $& $+1 $ & $-1 $  \\
$\sigma_{11}'$ & $(0^5;\frac{-1}{2}\,\frac{-1}{2}\,\frac{-1}{2})(0^5;\frac{3}{4}\,\frac{-1}{4}\,\frac{-1}{2})'$&$0$ &$ 4(1_1+1_3,1_{\bar{1}}+ 1_{\bar{3}} )$& $  4\cdot \one_{0}(T_3)$ &$-9$& $-6$ & $-6$ & $-6$ & $+12$ & $-9$ & $-3$ & $\frac{-30}{7}$ &  $-1  $& $+1 $ & $+1 $  \\
   $\sigma_{12}$ & $(0^5;\frac{-1}{2}\,\frac{1}{2}\,\frac{1}{2}\,)(0^5;\frac{3}{4}\,\frac{-1}{4}\,\frac{-1}{2})'$&$\frac{+1}{3}$ & $2(1_1+1_3,1_{\bar{1}}+ 1_{\bar{3}} )$ &$ 2\cdot \one_{0}(T_3)$  &$+3$& $-6$ & $+6$ & $+6$ & $+12$ & $-9$ & $-3$ & $\frac{+40}{7}$ &  $+1  $& $+1 $ & $-1 $  \\
   $\sigma_{12}'$ & $(0^5;\frac{-1}{2}\,\frac{1}{2}\,\frac{1}{2}\,)(0^5;\frac{3}{4}\,\frac{-1}{4}\,\frac{-1}{2})'$&$\frac{+2}{3}$ & $2(1_1+1_3,1_{\bar{1}}+ 1_{\bar{3}} )$ &$ 2\cdot  \one_{0}(T_3)$  &$+3$& $-6$ & $+6$ & $+6$ & $+12$ & $-9$ & $-3$ & $\frac{+40}{7}$ &  $-1  $& $+1 $ & $+1 $  \\
  $\sigma_{13}$ & $(0^5;\frac{ 1}{2}\,\frac{1}{2}\,\frac{-1}{2}\,)(0^5;\frac{-1}{4}\,\frac{3}{4}\,\frac{-1}{2})'$&$\frac{+1}{3}$ &$2(1_1+1_3,1_{\bar{1}}+ 1_{\bar{3}} )$ &$2\cdot \one_{0}(T_3)$  &$+3$& $+6$ & $+6$ & $-6$ & $-12$ & $-9$ & $-3$ & $\frac{+124}{7}$ &  $+1  $& $+1 $ & $-1 $  \\
   $\sigma_{13}'$ & $(0^5;\frac{ 1}{2}\,\frac{1}{2}\,\frac{-1}{2}\,)(0^5;\frac{-1}{4}\,\frac{3}{4}\,\frac{-1}{2})'$&$\frac{+2}{3}$ &$2(1_1+1_3,1_{\bar{1}}+ 1_{\bar{3}} )$ &$ 2\cdot  \one_{0}(T_3)$  &$+3$& $+6$ & $+6$ & $-6$ & $-12$ & $-9$ & $-3$ & $\frac{+124}{7}$ &  $-1  $& $+1 $ & $+1 $  \\
  $\sigma_{14}$ & $(0^5;\frac{ 1}{2}\,\frac{1}{2}\,\frac{-1}{2}\,)(0^5;\frac{-1}{4}\,\frac{-1}{4}\,\frac{1}{2})'$&$\frac{+2}{3}$ &$2(1_{\bar{1}})+1( 1_{\bar{3}} )$ &$ 3\cdot  \one_{0}(T_3)$  &$+7$& $+6$ & $+6$ & $-6$ & $0$ & $+9$ & $+3$ & $\frac{+58}{7}$ &  $-1  $& $+1 $ & $+1 $  \\
$\sigma_{15}$  & $(0^5;\frac{-1}{2}\,\frac{-1}{2}\,\frac{-1}{2})(0^5;\frac{+3}{4}\,\frac{-1}{4}\,\frac{-1}{2})'$&$\frac{+2}{3}$ &$2(1_1+1_3,1_{\bar{1}}+ 1_{\bar{3}} )$& $ 2\cdot\one_{0}(T_3)$&$-9$& $-6$ & $-6$ & $-6$ & $+12$ & $-9$ & $-3$ & $\frac{-30}{7}$ &   $+1  $& $+1 $ & $-1 $  \\
$\sigma_{15}'$  & $(0^5;\frac{-1}{2}\,\frac{-1}{2}\,\frac{-1}{2})(0^5;\frac{+3}{4}\,\frac{-1}{4}\,\frac{-1}{2})'$&$0$ &$2(1_1+1_3,1_{\bar{1}}+ 1_{\bar{3}} )$& $  4\cdot \one_{0}(T_3)$  &$-9$& $-6$ & $-6$ & $-6$ & $+12$ & $-9$ & $-3$ & $\frac{-30}{7}$ &   $-1  $& $+1 $ & $+1 $  \\
  $\sigma_{16}$  & $(0^5;\frac{-1}{2}\,\frac{+1}{2}\,\frac{+1}{2})(0^5;\frac{+3}{4}\,\frac{-1}{4}\,\frac{-1}{2})'$&$\frac{+1}{3}$ & $2(1_1+1_3,1_{\bar{1}}+ 1_{\bar{3}} )$ &$ 2\cdot \one_{0}(T_3)$   &$+3$& $-6$ & $+6$ & $+6$ & $+12$ & $-9$ & $-3$ & $\frac{+40}{7}$ &  $+1  $& $+1 $ & $-1 $  \\
 $\sigma_{16}'$  & $(0^5;\frac{-1}{2}\,\frac{+1}{2}\,\frac{+1}{2})(0^5;\frac{+3}{4}\,\frac{-1}{4}\,\frac{-1}{2})'$&$\frac{+2}{3}$ & $2(1_1+1_3,1_{\bar{1}}+ 1_{\bar{3}} )$ &$ 2\cdot  \one_{0}(T_3)$   &$+3$& $-6$ & $+6$ & $+6$ & $+12$ & $-9$ & $-3$ & $\frac{+40}{7}$ &  $-1  $& $+1 $ & $+1 $  \\
$\sigma_{17}$ & $(0^5;\frac{+1}{2}\,\frac{+1}{2}\,\frac{-1}{2})(0^5;\frac{-1}{4}\,\frac{+3}{4}\,\frac{-1}{2})'$&$\frac{+1}{3}$ &$2(1_1+1_3,1_{\bar{1}}+ 1_{\bar{3}} )$ &$ 2\cdot \one_{0}(T_3)$   &$+3$& $+6$ & $+6$ & $-6$ & $-12$ & $-9$ & $-3$ & $\frac{+124}{7}$ &   $+1  $& $+1 $ & $-1 $  \\
$\sigma_{17}'$ & $(0^5;\frac{+1}{2}\,\frac{+1}{2}\,\frac{-1}{2})(0^5;\frac{-1}{4}\,\frac{+3}{4}\,\frac{-1}{2})'$&$\frac{+2}{3}$ &$2(1_1+1_3,1_{\bar{1}}+ 1_{\bar{3}} )$ &$ 2\cdot \one_{0}(T_3)$  &$+3$& $+6$ & $+6$ & $-6$ & $-12$ & $-9$ & $-3$ & $\frac{+124}{7}$ &   $-1  $& $+1 $ & $+1 $  \\
  $\sigma_{18}$  & $(0^5;\frac{ 1}{2}\,\frac{+1}{2}\,\frac{-1}{2})(0^5;\frac{+3}{4}\,\frac{-1}{4}\,\frac{-1}{2})'$&$\frac{+2}{3}$ & $2(1_{\bar{1}})+1( 1_{\bar{3}} )$ &$ 2\cdot  \one_{0}(T_3)$   &$+7$& $+6$ & $+6$ & $-6$ & $0$ & $+9$ & $+3$ & $\frac{+58}{7}$ &  $-1  $& $+1 $ & $+1 $  \\
$\sigma_{21}$  & $(0^5;\frac{-1}{6}\,\frac{-1}{6}\,\frac{-1}{6})(0^5;\frac{1}{4}\,\frac{1}{4}\,\frac{-1}{2})'$&$0$ &$1(1_{\bar1})$& $ \one_{0}(T_{1}^0)$ &$-3$& $-2$ & $-2$ & $-2$ & $0$ & $+9$ & $+3$ & $\frac{+12}{7}$ & $-1$& $-1$ & $-1$  \\ 
$\sigma_{22}$  & $(0^5;\frac{-5}{6}\,\frac{1}{6}\,\frac{1}{6})(0^5;\frac{1}{4}\,\frac{1}{4}\,\frac{-1}{2})'$&$0$ &$1(1_{\bar1}+1_3)$& $ \one_{0}(T_{5}^0)$ &$+1$& $-10$ & $+2$ & $+2$ & $0$ & $+9$ & $+3$ & $\frac{-2}{7}$ & $-1$& $+1$ & $+1$   \\
$\sigma_{23}$  & $(0^5;\frac{1}{6}\,\frac{-5}{6}\,\frac{1}{6})(0^5;\frac{1}{4}\,\frac{1}{4}\,\frac{-1}{2})'$&$0$ &$1(1_{\bar1}+1_3)$& $ \one_{0}(T_{5}^0)$ &$+1$& $-10$ & $+2$ & $+2$ & $0$ & $+9$ & $+3$ & $\frac{-44}{7}$ & $-1$& $+1$ & $+1$   \\
$\sigma_{24}$  & $(0^5;\frac{1}{6}\,\frac{1}{6}\,\frac{-5}{6})(0^5;\frac{1}{4}\,\frac{1}{4}\,\frac{-1}{2})'$&$0$ &$1(1_{\bar1}+1_3)$& $ \one_{0}(T_{5}^0)$ &$+1$& $-10$ & $+2$ & $+2$ & $0$ & $+9$ & $+3$ & $\frac{+82}{7}$ & $-1$& $+1$ & $+1$   \\[0.2em]
  \botrule
\end{tabular} 
\end{center}
\caption{U(1) charges of   L-handed neutral scalars (but $\sigma_{7,8}$ for R-handed).   We kept up to one oscillators represented in $(N^L)_j$ meaning  {\it Number of resulting fields}$({\rm [number~ of~ oscillating~ mode]}_{\rm torus~ of~ oscillating~ mode})$. For example, $n(1_{\bar{1}})$ means that there results $n$ multiplicities with one oscillator with phase $\frac{-5}{12}$.  For $Q_{18,20,22}$ charges, here we listed only those of L-handed fields, participating in the Yukawa couplings.   }\label{tab:Scalars} 
\end{table}

\subsection{$\tenb_{-1}+\ten_{+1}$ needed for spontaneous breaking of \flip}
  In   $T_3$ and $T_9$, we have
\dis{
\tilde{\chi}(\theta^3,\theta^j)=\left\{\begin{array}{lrrrrrrrrrrrr} 
j=&0,&1,&2,&3,&4,&5,&6,&7,&8,&9,&10,&11 \\
&4,&1,&1,&4,&1,&1,&4,&1,&1,&4,&1,&1 \end{array}\right. \label{eq:MultT3}
} 
 In   $T^3$, we have  from (\ref{ThetaModel}),
\dis{
\Theta_3& =\sum_{i} (N^L_i-N_i^R)\hat{\phi}_i - \tilde{s}  \cdot\phi+P\cdot V. \label{eq:Theta3}
}
For $\Sigma_1^*$ with $P=(\underline{+++--};+++)(0^5;-1,-1,+2)'$ and $\Sigma_2$ with $P=(\underline{++---};---)(0^5;+1,+1,-2)'$, $P\cdot V$ is $+\frac14$ and $-\frac14$, respectively.  Without oscillators, the masslessness condition is not satisfied. For $s=( \oplus\textrm{ or }\ominus;-,\pm,-)$,   we obtain the following multiplicities for massless  $\Sigma_1^*$ and $\Sigma_2$,
\dis{
 \begin{array}{cccccc} 
 {s}  & (N^L_i-N^R_i)\hat{\phi}_i,  & \tilde{s}  \cdot\phi,&\Theta_3, &{\rm Multiplicity},&P\cdot V\\[0.3em]
 (\oplus|-+-):&\frac{+5}{12 }\,({\rm torus=}1), &\frac{-1}{12}, & 0  , & 2,&\frac{+1}{4} (\Sigma_1^*)\\[0.3em] 
 (\ominus|---):&\frac{-1}{12 }\,({\rm torus=}\bar{3}), &\frac{-5}{12}, &   \frac{+4}{12} , & 1,&\frac{-1}{4}(\Sigma_2) \\[0.3em] 
  \end{array}  \label{eq:SigmasA}
}
 \dis{
 \begin{array}{cccccc} 
 {s}  & (N^L_i-N^R_i)\hat{\phi}_i,  & \tilde{s}  \cdot\phi,&\Theta_3, &{\rm Multiplicity},&P\cdot V\\[0.3em]
 (\oplus|-+-):&\frac{+1}{12 }\,({\rm torus=}3), &\frac{-1}{12}, &  \frac{+8}{12} , &1, &\frac{+1}{4} (\Sigma_1^*)\\[0.3em] 
 (\ominus|---):&\frac{-5}{12 }\,({\rm torus=}\bar{1}), &\frac{-5}{12}, &   0 , & 2, &\frac{-1}{4}(\Sigma_2) \\[0.3em] 
  \end{array}  \label{eq:SigmasB}
}
For L-handed fields, we obtain $Q_{18,20,22}=(-1,+1,-1)$ and $(-1,-1,-1)$. The L-handed field multiplicities are from the oscillators of the 1st and 3rd tori, $\hat{\phi}_i=\pm\frac{5}{12},\pm\frac{1}{12}$ as shown in Eqs. (\ref{eq:SigmasA}) and (\ref{eq:SigmasB}). These are explicitly shown in Table II, including the respective multiplicities and charges $Q_{18,20 ,22}$.
     
\subsection{Neutral \flip~singlets}

U(1) charges of  all neutral \flip~singlets $\sigma$'s are listed in Table II.   For a few neutral \flip~singlets,   we present the explicit calculation   in Appendix \ref{appA}. In Table II, we list
U(1) charges of $\sigma$ type singlets.  Those appearing with oscillators form vectorlike representations out of which we kept only L-handed fields because R-fields would  appear with more  mass suppression factors in the Yukawa couplings of the SM fields.  We kept up to one oscillators allowed in Table \ref{tab:OsciilatingModes} presented in Appendix A.   For $Q_{18,20,22}$ charges, we listed only those of L-handed fields. We need some singlets carrying
\dis{
\Phi=-\frac{1}{4}
}
to cancel  all possible phases in the superpotential. But Table II does not include such a field, and we must consider 
$[\sigma_i(\frac{+1}{4})]^*$ to make a phase invariant combination by providing $\Phi=-\frac{1}{4}$, which will appear in Sec. \ref{sec:Yukawa}.

\section{Yukawa couplings}\label{sec:Yukawa}

\subsection{$\mu$ problem}\label{subsec:mu}
We find that there remain two pairs of L-handed $H_{uL}$ and $H_{dL}$ in $T_6$.  There exists  a superpotential term via $T_6 T_6$ where $T_6$ is a field appearing in the twisted sector $T_6$,   if the condition in Subsec. \ref{subsec:Selection} is satisfied. So, we expect a $\mu$ term at the GUT scale,
\dis{
W= -\mu_{\rm GUT}   H_{uL}^i  H_{dL}^j, ~\textrm{ for~} i,j=\{a,b\}. \label{eq:muTerm}
} 
With two $H_{uL}$ and two $H_{dL}$ of  Eq. (\ref{eq:HiggsT6}),  the condition of  Subsec. \ref{subsec:Selection} is not satisfied since the phase $\frac{4\pi}{3}$ of $H_{uL}^i  H_{dL}^j$ does not allow   Eq. (\ref{eq:muTerm}). This is because it does not carry two units of $2\pi$, needed for a superpotential term. The basic reason is that  the orbifold contains fixed points divisible by 3.  In this regard, we note that  $\Z_3$ orbifold has 27 fixed points, forbidding dimension-3  $\mu$   term  as   firstly shown in  \cite{Casas93}. $\Z_{12-I}$\,contains twisted sectors  where the number of fixed points is 3.  In $\Z_{12-I}$\,and $\Z_{6-I}$, Higgs doublets in  the sector $T_{N/2}(N=12,\textrm{ or } 6)$ form   vector-like representations of  massles L-handed fields. In $\Z_{12-I}$, these vector-like representations do not form a $\mu$ term because of the above comment on \UoR\,charge condition. $\Z_{6-II}$ has 12 fixed points and  forbidding the dimension-3 $\mu$   term  may be possible here also.  

Two pairs surviving from the dimension-3 couplings couple to GUT scale VEVs by high dimensional operators. In this case, since two pairs are just from the phase condition on $\Theta_6=\frac{1}{3}$ of $H_u^i\,(i=1,2)$ and $H_d^j\,(j=1,2)$ in Eq. (\ref{eq:AllHs}), these two are not distinguished. So, if   couplings of the $2\times 2~\mu$-matrix are democratic,
\dis{
\begin{pmatrix} \mu_{\rm GUT}& \mu_{\rm GUT}
\\    \mu_{\rm GUT}& \mu_{\rm GUT}
\end{pmatrix},\label{eq:muMatrix}
}
there remains only one light pair. The heavy pair obtains the $\mu$-term $2\mu_{\rm GUT}$. This is a remarkable result. A possible Yukawa coupling leading to (\ref{eq:muMatrix}) arises in a dimension-4 superpotential,
\dis{
W_\mu\propto \frac{1}{\tilde{M}^2}H_u( T_6,\frac13) H_d( T_6,\frac13)\cdot \left[\Sigma_1^*(T_3,0) \Sigma_2(T_3,\frac13)\right]\cdot\left\{\sigma_i\right\}
\label{eq:muGUT}
}
where $\tilde{M}$ is a string/GUT scale mass parameter and the multiplicity of $\Sigma_1^*$ and  $\Sigma_2$ with the phase 0 is 2 and  the multiplicity with the phase $\frac13$ is 1. 
A $\sigma_{i}$ is attached to make the \UoR~charge 2 modulo 4. Since the VEV of $\sigma_{i}(i=2,3,4)$ breaks $\Z_{4R}$ symmetry, $\tilde{M}$ and $\mu_{\rm GUT}$ are constrained such that the dimension-5 proton decay operator   is sufficiently suppressed. We note that $\mu_{\rm GUT}$ in  (\ref{eq:muMatrix}) is of order $ |\langle \Sigma_1^* \Sigma_2\sigma_2\rangle|/\tilde{M}^2$ where $\langle \Sigma_1^*\rangle=\langle \Sigma_2\rangle$ at a GUT scale are needed to break the \flip~to the SM. If we take $|\langle \Sigma_2\rangle|\sim  \tilde{M}\simeq 10^{17}\gev$, the scale   $\mu_{\rm GUT}$ is about $\langle \sigma_2\rangle $ where $\Z_{4R}$ is broken.

If we take the democratic form Eq. (\ref{eq:muMatrix}), the massless component of two $H_u$'s, $H_{ua} $ and $H_{ub} $,   is
\dis{
(H_u)_{\rm SM}=\frac{1}{\sqrt2}\left( H_{ua}  -H_{ub} \right).
}

\subsection{Vectorlike exotics}

$\Z_{4R}$ is broken at an intermediate scale by singlet VEVs of $Q_R=2$ modulo 4, and
all vector-like exotics   would obtain masses at the intermediate scale.
The gauge coupling unification toward the low energy value of $\sin^2\theta_W\simeq 0.231$ \cite{KimRMP81,CorfuEW18,CMSew}   can be studied for all the  intermediate scale masses  of these vector-like exotics. 

\subsection{Negative masses}
\dis{
2M_L^2&= (P+kV_f)^2+2  \tilde{c}_k ,\\[0.2em]
2M_R^2&=s_0^2 +(\tilde{s}+k\phi )^2 +2 c_k .
}

For the SM masses, we need some \flip~singlets developing GUT/string scale VEVs. There is no \flip~singlets in the untwisted sector $U$.  So,  some tachyonic scarars may be needed in the twisted sectors, \ie at the origin some scalar mass must be negative. This condition for the right movers in the $k$-th twisted sector accompanies with the condition $M_L^2=M_R^2$, 
\dis{
\textrm{Right mover: } 2M_R^2&=(\tilde{s}+k\phi )^2<-2 c_k -\frac14=
\left\{ \begin{array}{l}\frac{5}{24}~\textrm{for }k=1,\\[0.3em]
\frac{1}{4}~\textrm{for }k=2,\\[0.3em]
\frac{3}{8}~\textrm{for }k=3,\\[0.3em]
\frac{1}{12}~\textrm{for }k=4,\\[0.3em]
\frac{5}{24}~\textrm{for }k=5,\\[0.3em]
\frac{1}{4}~\textrm{for }k=6.
\end{array} \right.\label{eq:negMassR}
}
For L-handed fields, we have 
\dis{
 \begin{array}{cc|c|c} \toprule
 {s}    & \tilde{s}\cdot\phi,&(\tilde{s}+k\phi )^2, &{\rm Check}~  M^2<0~{\rm for}~k=\\ \hline\\[-1.0em]
 (\ominus|---): &\frac{-5}{12}, &\frac{5}{24}, \frac{1}{ 4} ,\frac{3}{8}, \frac{ 12}{144} ,\frac{5}{24}, \frac{1}{4}   & 1(\times),2(\times),3(\times),4(\times),5(\times),6(\times) 
 \\[0.3em] 
 (\ominus|-++):  &0,   &  &1(\times),2(\times),3(\times),4(\times),5(\times),6(\times) 
 \\[0.3em] 
(\ominus|+-+): &\frac{+1}{12},   &  & 1(\times),2(\times),3(\times),4(\times),5(\times),6(\times) 
\\[0.3em] 
 (\ominus|--+):  &\frac{+4}{12},   &   & 1(\times),2(\times),3(\times),4(\times),5(\times),6(\times)   \\[0.3em] \hline
  \end{array}  
}
We checked the first row to see whether some mass is negative, but there is no negative mass states which are symbolically shown  with $\times$. The next three rows have larger values of $(\tilde{s}+k\phi )^2$ and again there  is no negative mass states. Overall, there is no negative mass states from string compactification. The needed VEVs must arise with appropriate Yukawa couplings.

\subsection{GUT breaking and $\Z_{4R}$}

Let us define U$_R$ charges  such that matter fields carry +1 unit in the following way,
 \dis{
 Q_R=\frac12(Q_1+Q_2+Q_3)+\frac{1}{6}(Q_5+Q_6)+2Q_{20},\label{eq:Rfinal}
}
which  are listed in Tables I and II. By giving VEVs to $Q_R=4$ \flip~field(s), we obtain the discrete symmetry  $\Z_4$. This is possible with the GUT breaking VEVs $\langle \Sigma_1^*\rangle=\langle \Sigma_2\rangle$. If any other $\sigma$ singlet, carrying $Q_R\ne 4$ modulo 4, develops a VEV then it will break  $\Z_4$.
  
\subsection{Top quark mass}
Top quark is in the $U$ sector. The selection rule of Subsec. \ref{subsec:Selection} requires the following coupling 
\dis{
\sim \frac{1}{\tilde{M}^2} t(U_3,0) t^c(U_3,0)H_u(T_6,\frac{1}{3})\Sigma_1^*(T_3,\frac{2}{3}) \Sigma_2(T_3 ,0),\label{eq:tMassForm}
}
where $\tilde{M}$ is some string/GUT scale and the 2nd numbers in the brackets are $\Theta_i$'s given in Tables I and II. $Q_R$ of $T_3$ fields (necessarily developing a GUT scale VEVs as required for breaking \flip) add up to 0 modulo 4. Thus, the total $Q_R$ of (\ref{eq:tMassForm}) is $+2$ which is the required \UoR~charge in the superpotential.   Then, the top quark mass is  
\dis{
m_{tt}\sim \langle H_{uL}\rangle\frac{ M_{10}^2}{\tilde{M}^2},\label{eq:tMass}
}
where $M_{10}=|\langle\Sigma_1^* \rangle|=|\langle\Sigma_2  \rangle| $.
The bottom quark mass arises similarly from
\dis{
 \propto b(U_3,0) b^c(U_3,0) H_d(T_6,\frac{1}{3})\Sigma_1^*(T_3,\frac{2}{3}) \Sigma_2(T_3 , 0).\label{eq:bMassForm}
}

\subsection{Proton decay problem}

The most dangerous operator for proton decay is the dimension-5 operators composed of matter fields,  $u,d, s,c$, $b,t, e,\mu,\tau, \nu_e, \nu_\mu,\nu_\tau$ in Table I,
\dis{
q^Iq^Jq^K\ell^L,\label{eq:pDecayDim5}
}
where $I,J,K,$ and $L$ are family indices.
As shown in Table I, the $\Z_{4R}$ quantum numbers are $-1$ for the SM fermions. A superpotential term is allowed if $Q_R=2$ modulo 4. Thus, the dimension-5 proton decay operator from the  superpotential (\ref{eq:pDecayDim5}) is not allowed.   
 
\subsection{Neutrino masses}
The neutrino mass operator is
\dis{
M^\nu_{33} &\propto\frac{1}{\tilde{M}_3^3}\int d^2\vartheta   \, \bar{\eta}_3(U_3,0)\bar{\eta}_3(U_3,0)  H_{uL}(T_6,\frac13) H_{uL}(T_6,\frac13)\Sigma_1^*(T_3,\frac23)\Sigma_1^*(T_3,\frac23) \\
M^\nu_{22}\propto &\frac{1}{\tilde{M}_2^4} \int d^2\vartheta  \,\bar{\eta}_2(T_4^0,\frac14)\bar{\eta}_2(T_4^0,\frac14)   H_{uL}(T_6,\frac13) H_{uL}(T_6,\frac13)\Sigma_1^*(T_3,\frac23)\Sigma_1^*(T_3,\frac23)  \sigma_5(T_6,\frac12)\\
 \label{eq:nuMass}
}
where $\tilde{M}_{3,2}$ are some GUT scale mass parameters, and we satisfied $Q_R=2$ above for $d^2\vartheta$ integration. 
Then, the above  masses are estimated as 
\dis{
M^\nu_{33}\sim \frac{v_{EW}^2  }{\tilde{M} _3} \frac{ M_{10}^2 }{\tilde{M}_3^2}, ~M^\nu_{22}\sim \frac{v_{EW}^2  }{\tilde{M}_2 }  \frac{ M_{10}^2 | \langle\sigma_5  \rangle |}{\tilde{M}_2^3}  .\label{eq:nuOrder}
}
Then, above neutrino  masses are of order $v_{EW}^2/\tilde{M}$ since the SM singlet VEVs, $ M_{10}$ and $ | \langle\sigma_5  \rangle |$ can be at the GUT scale without breaking $\Z_{4R}$.

 To obtain mixing between $U_3$ and $T_4^0$ neutrinos, we need $d^2\vartheta d^2\bar{\vartheta}$ integration, \ie require  $Q_R=0$ modulo 4 for $d^2\vartheta d^2\bar{\vartheta}$ integration,
\dis{
M_{32}^\nu, M^\nu_{31}\propto \frac{1}{ \tilde{M}^5_m }\int d^2\vartheta d^2\bar{\vartheta} \, \bar{\eta}_3(U_3,0)\bar{\eta}_{2,1}(T_4^0,\frac{1}{4})    H_{uL}(T_6,\frac13) H_{uL}(T_6,\frac13)\Sigma_1^*(T_3,\frac23)\Sigma_1^*(T_3,0)  \cdot  \sigma_1(T_4^0,\frac12)^* .
}
Then, the above  mass mixing is estimated as 
\dis{
M^\nu_{13,23}\sim  \frac{v_{EW}^2  }{\tilde{M}_m }  \frac{ M_{10}^2 | \langle\sigma_1  \rangle |}{\tilde{M}_m^3} .\label{eq:nu23mixing}
}
 $\Sigma_1^*$ and $\Sigma_2$  can have the GUT scale VEVs because all of them carry $Q_R=4$, but  and $|\sigma_1|\ll \tilde{M}_m$ because it breaks $\Z_{4R}$. Depending on the ratio ${ M_{10}^2 | \langle\sigma_1  \rangle |}/{\tilde{M}_m^3} $, the mixing masses can be tuned.  
 
 Comparing   $M^\nu_{11,22,31,32}$ and $M^\nu_{33}$, 
\dis{
\frac{M^\nu_{11},M^\nu_{22}}{M^\nu_{33}}\approx \left|\frac{\sigma_{5} }{\tilde{M}} \right|,
\frac{M^\nu_{31},M^\nu_{32}}{M^\nu_{33}}\approx \left|\frac{\sigma_{1} }{\tilde{M}} \right|,
}
we note that the neutrino mass hierarchy favors the normal hierarchy (in the sense that $\nu_\tau$ is the heaviest) if the VEVs of $\sigma$ singlets are comparably small, $|\sigma_{1}|,|\sigma_{5}| \ll \tilde{M}$.

\subsection{Mass matrices, and CKM and PMNS mixing angles}

Mass matrices obtained in the weak basis are diagonalized to give the CKM matrix in the quark sector and the PMNS matrix in the lepton sector. The Yukawa couplings allowed by the $\Z_{4R}$ quantum numbers, shown in Tables I and II, dictate the forms of mass matrices in the weak basis. Fitting to the observed CKM angles in some detail are presented in a separate paper \cite{Jeong18}.  For the PMNS matrix, the observed data are not accurate
 enough to analyze it now. 
 
 \section{The vacuum structure}\label{sec:Vacuum}
 
 In this section, our main interest is how the vacuum at the GUT scale, leading to the $\Z_{4R}$ discrete symmetry \cite{Lee11}, is realized in our scheme. The following U(1)$_R$ quantum numbers are determined if $\Sigma_1^*, \Sigma_2, \sigma_{5}, \sigma_{6}, \sigma_{7}$, and $\sigma_{8}$ develop GUT scale VEVs. All the other fields are not required to have a GUT scale VEV. Then, there remains a degeneracy which we remove  by requiring a simple form for $Q_R$. Let  us start by parametrizing the U(1)$_R$ charge, without including the anomaly free $Q_4$, as
 \dis{
 Q_R=x_1 Q_1 +x_2(Q_2+a_3  Q_3) +x_5Q_5+x_6Q_6 +x_{20}(k_{18} Q_{18}+ Q_{20}+k_{22} Q_{22}).  \label{eq:Rfpara}
 }
 To break the \flip~down to a supersymmetric SM, $\Sigma_1^*$ and $\Sigma_2$ must develop the same magnitude VEV. Therefore, the contributions from the KK sector must be mutually exactly opposite (by considering effective D-terms) for $\Sigma_1^*$ and $\Sigma_2$. This condition is on the gauge charges and hence, toward SUSY below the GUT scale, we must require $k_{18}=k_{22}=0$.  Nonzero VEVs of $\sigma_5$ and  $\sigma_6$, leading to the same discrete charge of $\sigma_5$ and  $\sigma_6$ (for $\sigma_7$ and  $\sigma_8$ also), gives a possibility $a_3=1$. If $x_1=x_2$, we have
 \dis{
 Q_R=x_1(Q_1+Q_2+Q_3)+x_5Q_5+x_6Q_6+x_{20}Q_{20}.\nonumber
 }   
As an illustration, let us try $x_{20}=2$. To have a $\Z_4$ subgroup from the VEVs of $\Sigma_1^*$ and $\Sigma_2$, from $Q_5,Q_6$ and $Q_{20}$ charges in Table II, we have $x_5Q_5+x_6Q_6=\pm 2$ for $\Sigma_1^*$ and $\Sigma_2$, respectively.  Then,  note that  $x_5Q_5+x_6Q_6=\pm 2$ or 0 in Tables I and II. For the matter fields of Table I to have an odd $Q_R$, we  fix $x_1=x_2 =\frac12$. Still, $x_5$ and $x_6$ are not determined. For an illustration, we can choose   $x_5=x_6=\frac16$ such that
 \dis{
 Q_R=\frac{1}{2}(Q_1+Q_2+Q_3)+\frac{1}{6}(Q_5+Q_6)+2Q_{20}. \label{eq:Rfinal}
 }
 In Tables I and II, $Q_R$ of Eq. (\ref{eq:Rfinal}) are presented. This illustration is realized if the following conditions are met for the vacuum from gauge symmetries: 
     \begin{itemize}
   \item[(a)] SUSY is realized below the GUT breaking scale.
   \item[(b)] There exist  VEVs for $\sigma_5$ and  $\sigma_6$, and also  for $\sigma_7$ and  $\sigma_8$. Also, there exists a VEV of $\sigma_1$.
   \item[(c)] $x_1=x_2=\frac{1}{4}x_{20}$. 
    \item[(d)] Realization of $\Z_{4R}$. 
 \end{itemize}
 Items (a) and (d) are what we want. Item (a) is automatically fulfilled in our construction because we obtained a SUSY flipped SU(5) from the orbifold construction \cite{Dixon2,Ibanez1}. Item (d) follows if  Items (b) and (c) are fulfilled. We can see it by choosing $x_{20}=2$, for which we obtain odd numbers for matter fields of Table I, and there is no fractional numbers in  Tables I and II.   Thus,   $\Z_{4R}$ is realized. In the remainder of this section, therefore, we discuss the points related to Items (b) and (c).
 
Item (b) requires showing that  $\sigma_5$ and  $\sigma_6$  and also $\sigma_7$ and  $\sigma_8$ develop VEVs. The BSM fields in Table II can have the following $\vartheta$-dependent gauge invariant terms in the superpotential,
\dis{
\Sigma_1^*(T_3,\frac23)\Sigma_2(T_3,\frac13),\Sigma_1^*(T_3,0)\Sigma_2(T_3,\frac13,0),  \sigma_5(T_6,\frac12)\bar{\sigma}_7(T_6,-\frac12),\\
\sigma_6(T_6,\frac12)\bar{\sigma}_8(T_6,-\frac12),
 \sigma_5(T_6,\frac12)\sigma_6(T_6,\frac12)\bar{\sigma}_7(T_6,-\frac12)  \bar{\sigma}_8(T_6,-\frac12),\nonumber
}
plus any combinations of gauge invariant field products in Table II, having $\sum_i\Theta_i=0$. Consider a superpotential constructed with the above quadratic combinations,
\dis{
W=&-m_1\Sigma_1^*\Sigma_2-m_2 \sigma_5\bar{\sigma}_7 
-m_2'\sigma_6\bar{\sigma}_8+ \frac{\lambda_1}{2M}( \Sigma_1^* \Sigma_2)^2   \\[0.5em]
+& \frac{\lambda_2}{2M}( \sigma_5  \bar{\sigma}_7 )^2 + \frac{\lambda_2'}{2M} ( \sigma_6   \bar{\sigma}_8 )^2  
+\frac{\lambda_2''}{M} \sigma_5 \sigma_6 \bar{\sigma}_7  \bar{\sigma}_8+ \frac{\lambda_3}{M}  \Sigma_1^* \Sigma_2  \sigma_5  \bar{\sigma}_7+ \frac{\lambda_3'}{M}  \Sigma_1^* \Sigma_2  \sigma_6  \bar{\sigma}_8, \label{eq:Weff}
}
where, for simplicit, we considered only one combination of $\Sigma_1^*\Sigma_2$. Since $\sigma_5,\sigma_6, \bar{\sigma}_7$ and $\bar{\sigma}_8$ appear in the same twisted sector, $T_6$, later we set for simplicity: $m_2=m_2',~\lambda_2=\lambda_2'=\lambda_2''$ and $\lambda_3=\lambda_3'$. The SUSY conditions require
\dis{
\delta \Sigma_1^* &:~-m_1\Sigma_2+\frac{\lambda_1}{M}\Sigma_2\Sigma_1^* \Sigma_2+  \frac{\lambda_5}{M}   \Sigma_2  \sigma_5  \bar{\sigma}_7+ \frac{\lambda_6}{M}    \Sigma_2  \sigma_6  \bar{\sigma}_8=0,\\
\delta \Sigma_2 &:~-m_1\Sigma_1^*+\frac{\lambda_1}{M} \Sigma_1^* \Sigma_2 \Sigma_1^*+  \frac{\lambda_5}{M}   \Sigma_1^*  \sigma_5  \bar{\sigma}_7+\frac{\lambda_6}{M} \Sigma_1^* \sigma_6 \bar{\sigma}_8=0,\\
\delta  \sigma_5&:~-m_2 \bar{\sigma}_7 + \frac{\lambda_2}{M}    \bar{\sigma}_7\sigma_5  \bar{\sigma}_7  +  \frac{\lambda_4}{M}   \sigma_6 \bar{\sigma}_7  \bar{\sigma}_8+ \frac{\lambda_5}{M}  \Sigma_1^* \Sigma_2   \bar{\sigma}_7=0,\\
\delta  \sigma_6&:~-m_3 \bar{\sigma}_8 + \frac{\lambda_3}{M} \bar{\sigma}_8\sigma_6 \bar{\sigma}_8+  \frac{\lambda_4}{M}   \sigma_5 \bar{\sigma}_7  \bar{\sigma}_8+ \frac{\lambda_6}{M}  \Sigma_1^* \Sigma_2   \bar{\sigma}_8=0,\\
\delta \bar{\sigma}_7&:~-m_2  {\sigma}_5 + \frac{\lambda_2}{M}\sigma_5  \bar{\sigma}_7{\sigma}_5  +  \frac{\lambda_4}{M}  {\sigma}_5 \sigma_6\bar{\sigma}_8+ \frac{\lambda_5}{M}  \Sigma_1^* \Sigma_2 \sigma_5=0,\\
\delta \bar{\sigma}_8&:~-m_3  {\sigma}_6+ \frac{\lambda_3}{M} \sigma_6 \bar{\sigma}_8 {\sigma}_6+  \frac{\lambda_4}{M}   \sigma_5\sigma_6 \bar{\sigma}_7+ \frac{\lambda_6}{M}  \Sigma_1^* \Sigma_2{\sigma}_6=0.
}
We choose the vacuum where all of the above fields develop VEVs, $V_i=\langle\sigma_i\rangle$ and $V_{10}= \langle\Sigma_1^*\rangle= \langle\Sigma_2\rangle$. In terms of $m_{1,2} $ and $\lambda_{1,2,3}$, we obtain two independent relations,
\dis{
 &-m_1 +\frac{\lambda_1}{M}V_{10}^2+  \frac{\lambda_3}{M} (V_5 V_7+V_6 V_8)=0,\\
&-m_2 + \frac{\lambda_3}{M}  V_{10}^2+ \frac{\lambda_2}{M}(V_5V_7  +   V_6 V_8)=0,
}
from which we conclude that  the singlets $\Sigma_1^*,\Sigma_2$, and $\sigma_i(i=5,6,7,8)$ develop GUT scale VEVs,
\dis{
V_{10}^2 =\frac{(\lambda_2 m_1-\lambda_3 m_2) }{\lambda_1\lambda_2-\lambda_3^2}M,\\
V_5V_7+V_6V_8 =\frac{(\lambda_3 m_1-\lambda_1 m_2) }{\lambda_3^2-\lambda_1\lambda_2}M.\label{eq:GUTvevs}
}

 We also need $\sigma_i(i=1,2,3,4)$ VEVs   which are much smaller than the GUT scale such that the B violating dimension-5 operators are sufficiently suppressed because  $\langle\sigma_i(i=1,2,3,4)\rangle$ would break $\Z_4$ down to $\Z_2$. These VEVs are considered to be a perturbation to the VEVs of (\ref{eq:GUTvevs}), and $\langle\sigma_i(i=1,2,3,4)\rangle$  would be close to $0$. Note that $\sigma$'s in Table II are not moduli, and there is no $\sigma_i$ with all gauge charges vanishing. Therefore, not to produce  runaway solutions of  $\sigma_i$,  the mass parameters in the numerator of renormalizable terms and in the denominater of non-renormalizable terms, leading to $\langle\sigma_i(i=1,2,3,4)\rangle$, are required to be of sub-GUT scale. Consider the following gauge and $\Theta$ invariant D-terms containing $\sigma_{i}(i=1,2,3,4)$ for $m_I, M_I \ll \tilde{M}$,
\dis{
-\frac12m_I\sum_{i=1}^4\int d^2\vartheta d^2\bar{\vartheta}\, \sigma_i\sigma_i^* ,~
\frac{\lambda_I}{2M_I^8}\int d^2\vartheta d^2\bar{\vartheta}\, \sigma_1^2\sigma_2^2\sigma_3^2\sigma_7^2 \sigma_8^2
}
where $\sigma_i$ carry $\vartheta$, and $\sigma_{7,8}$ and  $\sigma_i^*$ carry $\bar\vartheta$.  One among $\sigma_{7}$ and $\sigma_{8}$ carries 2 KK windings and the others carry 1 KK winding such that U(1)$_{20}$ invariance is satisfied. Then, $V$ contains the following
\dis{
V\ni  -\frac12 m_I^2\sum_{i=1}^4|\sigma_i|^2 +\frac{\lambda_I}{2M_I^8}\left[\sigma_1(\sigma_1^*)^2 \sigma_2^2 \sigma_3^2\sigma_7^2 \sigma_8(\sigma_8^*)^2+\cdots\right] +|-m_2  {\sigma}_6+ \frac{\lambda_2}{M} \sigma_6 \bar{\sigma}_8 {\sigma}_6+  \frac{\lambda_2}{M}   \sigma_5\sigma_6 \bar{\sigma}_7+ \frac{\lambda_3}{M}  \Sigma_1^* \Sigma_2{\sigma}_6|^2,
}
leading to
\dis{
\frac{dV}{d\sigma_1^*}=0\to\frac{\lambda_I}{M_I^8} \left[\sigma_1\sigma_1^*  \sigma_2^2 \sigma_3^2\sigma_7^2 \sigma_8(\sigma_8^*)^2+\cdots\right]   = m_I^2\sigma_1,
}
or
\dis{
   \sigma_1^*  =\frac{m_I^2 M_I^8}{\lambda_I \Big( \sigma_2^2 \sigma_3^2\sigma_7^2 \sigma_8(\sigma_8^*)^2+\cdots\Big)}   .\label{eq:IntV}
}
For $\sigma_5=\sigma_6=\sigma_8= \bar{\sigma}_7=V_5$ and $\sigma_1=\sigma_2=\sigma_3=\sigma_4=V_I$,
\dis{
V_I =\left(\frac{m_I^2 M_I^3}{\lambda_I}\right)^{1/5}\frac{M_I}{V_5}   .
}
where we neglected $\cdots$ in Eq. (\ref{eq:IntV}). If $\sqrt{m_IM_I}\approx 10^{-3} V_5$, then we obtain an intermediate scale $V_I\approx (\lambda_I)^{1/5}10^{-6}V_5$. This can be a kind of models realizing the scale of the ``very light'' axion in supergravity models \cite{Kim84W}.
So, we conclude that the vacuum, satisfying Item (b), can be realized. In addition, if the solution for the $\mu$ term is realized {\it \`a la} \cite{KimNilles84}, then the electroweak scale may be obtained along our vacuum direction.
 
  Item (c) requires showing that the quantum numbers of $x_2$ is $\frac34$ times $(3x_5+x_6)$. Referring to the six U(1) gauge quantum numbers of Tables I and II, integers are possible if $x_1, x_2, x_2a_3, x_5, x_6$, and $x_{20}$ are integer multiples of $\frac12, \frac12, \frac12,\frac13,\frac13$, and $1$, respectively. Item (a) requires the relation,
\dis{
 3x_5+x_6=\frac{x_{20}}{3} .
}
If we choose $x_{20}=2x$, we obtain $x_2=\frac12x$. So far, there remain  two degrees of freedom, \ie   arbitrary $x_1$ and  $x_6(=\frac{x_{20}}{3}-3x_5)$. Requiring a VEV for $\sigma_1$, we note that  $\langle \sigma_1\rangle$ breaks one U(1) gauge symmetry: U(1)$_{\sigma_1}$.  The breaking direction of U(1)$_{\sigma_1}$ should not affect other gauge symmetries. Thus, we require $x_1=x_2=x_2a_3$ in Eq. (\ref{eq:Rfpara}) because $Q_1, Q_2,$ and $Q_3$ quantum numbers of $\sigma_1$ in Table II are the same. This proves Item (c).

There remains one free parameter $x_6$. We cannot determine this parameter by VEVs of scalars. To give a smaller number for coefficients, we choose $x_5=x_6$, and the  result is Eq. (\ref{eq:Rfinal}) and Tables I and II.
   
\section{Conclusion}\label{sec:Conclusion} 

We obtained an R-parity as a discrete subgroup of \UoR~global symmetry of  \UEE$\times$\UKK~where \UEE~is the part from $\EE8$ and \UKK~is the part generated by $Q_{18},Q_{20}$, and $Q_{22}$.  We checked that the needed VEVs toward flavor mixing, the $\mu$ term, the neutrino mass operators, and forbidding dangerous dimension-5 $\Delta B\ne 0$ operators, are consistent with the \UoR~direction.  It has been possible because the number of $Q_{R}=4$ (modulo 4) fields of Table II are enough to render the needed operators.
One more interesting feature is that the SM quarks and leptons of Table I carrying $Q_{R}=-1$ (modulo 4) is not enough  by itself  to cancel the unwanted    dimension-5 $\Delta B\ne 0$ operators.  But the oringinal \UoR~charge helps to forbid these unwanted terms because the origin of $\Z_{4R}$ symmetry in the ultra-violet completed theory is the global \UoR~which forbid the unwanted  dimension-5 $\Delta B\ne 0$ operators. 

\begin{appendix}

\section{\UKK~charges of neutral singlets}\label{appA}

The definition of shift $\phi$ and the number of fixed points $\chi$ in $\Z_N$ orbifolds are presented in Table \ref{tab:OsciilatingModes} together with the allowed oscillating modes. The smallest number of fixed points is 3 which are possible in $\Z_{12-I}$ and $\Z_{6-I}$. Among these, we studied $\Z_{12-I}$ which allows more possibilities of Yukawa couplings.

For the massless modes, the phase determining the multiplicity is given in Eq. (\ref{PhaseMult}). The massless modes relevant for the Higgs mechanism in our model is for just $V$, \ie we do not use the fields from Wilson line added shifts. So, we set $V_f=V$ in   Eq.   (\ref{PhaseMult}),
\dis{
\Theta_k^0&=\sum_{i} (N_i^L-N_i^R)\hat{\phi}_i +P\cdot V - {s} \cdot \phi +\frac{k}{2}(V^2-\phi^2)\\&=\sum_{i} (N_i^L-N_i^R)\hat{\phi}_i +P\cdot V - {s} \cdot \phi +\frac{k}{12} .\label{Theta0}
} 

\dis{
V^2=\frac{11}{24},~\phi^2=\frac{7}{24},~\frac{V^2-\phi^2}{2}=\frac{1}{12},
}

In the main text, we illustrated the U(1) charge calculation for \flip~non-singlets. 
In this Appendix, we illustrate the calculational methods of massless spectrum and   the  U(1) charges  explicitly.  Both the left-mover and right-mover massless states satisfy
\dis{
&\textrm{Left mover: } M_L^2=\frac{(P+k V_f)^2}{2} + \tilde{c}_k=0,\\[0.4em]
&\textrm{Right mover: } M_R^2= \frac{({s}+k \phi)^2}{2}+  {c}_k=0,~
 s=(\oplus\textrm{ or }\ominus; \tilde{s})  
 . \label{eq:MassLess}
}
In Eq. (\ref{eq:MassLess}), $\oplus$ or $\ominus$ is chosen such that the total number of minus signs is even.

\subsection{\bf Two families from $T_4^0$}\label{app:Matter}
 For matter fields in $T_4^0$ without oscillators, we insert $P\cdot V=-\frac14$ for $k=4$ in Eq. (\ref{Theta0}),  
\dis{
\Theta_4^0({\rm matter})& =  - \tilde{s} \cdot \phi +\frac{1}{12} . 
} 
Note that $4\phi$ is $(\frac{20}{12},\frac{16}{12},\frac{4}{12})\to (\frac{2}{3},\frac{1}{3},\frac{1}{3})$ where we must use the entries in the region $[0,1)$. It is like the shift in $\Z_3$ orbifold. With this $4\phi$, the masslessness condition is $s_0^2+(\tilde{s}+k\phi)^2=-2c=\frac13$, \ie $(\tilde{s}+k\phi)^2=\frac{1}{12}$, which is satisfied by $\tilde{s}=(---)$. So, we choose $s=(\ominus;\tilde{s})$, \ie it is L-handed, and obtain $\Theta_4^0({\rm matter})=\frac{1}{2}$. With the following multiplicity contribution,
\dis{
\tilde{\chi}(\theta^4,\theta^j)=\left\{\begin{array}{lrrrrrrrrrrrr} 
j=&0,&1,&2,&3,&4,&5,&6,&7,&8,&9,&10,&11 \\
&27,&3,&3,&3,&27,&3,&3,&3,&27,&3,&3,&3\,,\end{array}\right.\label{eq:Chi4}
}
we obtain ${\cal P}=2$,  and the charges $Q_{18}, Q_{20}$ and $Q_{22}$ of matter  from $T_4^0$ are $-1,-1$, and $-1$, respectively, which are listed in Table I.

For Higgs fields $H_u$ with $P=(\underline{1\,0\,0\,0\,0};1\,1\,1)(0^8)'$ in $T_4^0$, we have $P\cdot V=-\frac12$. For $k=4$ in Eq. (\ref{Theta0}),  
\dis{
\Theta_4^0({\rm Higgs})& =  - \tilde{s} \cdot \phi +\frac{-2}{12} . 
} 
Since $4\phi$ is $(\frac{2}{3},\frac{1}{3},\frac{1}{3})$, the masslessness condition is the same as above,   $s=(\ominus;---)$, \ie it is left-handed(L-handed), and we obtain $\Theta_4^0({\rm Higgs})=\frac{1}{4}$. Again, we obtain ${\cal P}=2$ using Eq. (\ref{eq:Chi4}),  and the charges $Q_{18}, Q_{20}$ and $Q_{22}$ of Higgs from $T_4^0$  are $-1,-1$, and $-1$, respectively. These were used for the Higgs fields in \cite{KimFlavor18}, but here we will not use these for the Higgs fields for breaking the SM.

\begin{table}[t!]
\begin{center}
\begin{tabular}{@{}|ccc|c|c| @{}} \toprule
  ~$\Z_N$~ & $\phi$&$\chi$&Allowed oscillating mode $N_i$ \\[0.1em] \colrule
 ~$\Z_{12-I}$~ &$(\frac{5}{12},\frac{4}{12},\frac{1}{12})$& $3$ & $1_{1},1_{\bar{1}},1_{3},1_{\bar{3}}$  \\[0.3em]
 ~$\Z_{12-II}$~ &$(\frac{6}{12},\frac{5}{12},\frac{1}{12})$& $4$ &~ $1_{2},1_{\bar{2}},1_{3},1_{\bar{3}},2_{2},2_{\bar{2}},2_{3},2_{\bar{3}},3_{2},3_{\bar{2}},3_{3},3_{\bar{3}}$~  \\[0.3em]
 ~$\Z_{8-I}$~ &$(\frac{3}{8},\frac{2}{8},\frac{1}{8})$& $4$ & $1_{2},1_{\bar{2}},1_{1+3},1_{\bar{1}+\bar{3}} $  \\[0.3em]
 ~$\Z_{8-II}$~ &$(\frac{4}{8},\frac{3}{8},\frac{1}{8})$& $8$ & $1_{1},1_{\bar{1}},1_{2+3},1_{\bar{2}+\bar{3}} $  \\[0.3em]
~$\Z_{7}$~ &$(\frac{3}{7},\frac{2}{7},\frac{1}{7})$& $7$ & $0$  \\[0.3em]
~$\Z_{6-1}$~ &$(\frac{2}{6},\frac{1}{6},\frac{1}{6})$& $3$ & $0$  \\[0.3em]
~$\Z_{6-II}$~ &$(\frac{3}{6},\frac{2}{6},\frac{1}{6})$& $12$ & $1_{1},1_{\bar{1}} $  \\[0.3em]
 ~$\Z_{4}$~ &$(\frac{2}{4},\frac{1}{4},\frac{1}{4})$& $16$ & $1_{1},1_{\bar{1}} $  \\[0.3em]
~$\Z_{3}$~ &$(\frac{2}{3},\frac{1}{3},\frac{1}{3})$& ~$27$~ & $0$  \\[0.3em]
  \botrule
\end{tabular} 
\end{center}
\caption{Allowed  mode $N_i$ for calculating $\Theta$.  }\label{tab:OsciilatingModes} 
\end{table}
\vskip 0.5cm
\subsection{\bf $\sigma_{1-4}$ from $T_4^0$}
For $T_4^0$, we have calculated above the chirality as $s=(\ominus;---)$. With $P=(0^8)(0^5;-1,-1,0)'$ for $ \sigma_{1}$, we have  $P\cdot V=\frac{-1}{2}, \Theta_4=\frac{+3}{12}$, and obtain ${\cal P}=2$ without an oscillator.   With $P=(0^5;0,+1,+1)(0^8)'$ for $ \sigma_{2}$, we have  $P\cdot V=\frac{-4}{12}$ and $\Theta_4=\frac{+5}{12}$. With the oscillator $1_{\bar{1}}$, we obtain ${\cal P}=3$. For $\sigma_{1,2,3,4}$ we have
 \dis{
Q_{18,20,22}=(-1,-1,-1).
}
 
\subsection{\bf $\sigma_{5-8}$ from $T_6$}
 For $T_6$, the multiplicity factor and the phase are given in Eqs. (\ref{eq:MultT6}) and  (\ref{Theta6}). The allowed chiralities are $s=(\oplus|-+-)$ and $ s=(\ominus|---)$ for $6\phi=(\frac12,0,\frac12)$.
For $\sigma_{5}$, we use $P=(0^5;+1,+2,+1)(0^5;0,-1,+1)'$ and obtain    $P\cdot V= \frac{-5}{12}$, and   massless fields arise without oscillators,
  \dis{
 \begin{array}{ccccc} 
 {s}  & N_i,  & \tilde{s}  \cdot \phi,&\Theta_6, &{\rm Multiplicity}\\[0.3em]
 (\oplus|-+-):&0, &\frac{-1}{12}, & \frac{+2}{12},&0\cdot\sigma_{ 5,6}\\[0.3em] 
 (\ominus|---):&0, & \frac{-5}{12}, & \frac{+6}{12},& 2\cdot\sigma_{5,6} 
  \end{array}  
}
 For $\sigma_{7}$, we use $P=(0^5;+1,0,+1)(0^5;0,-1,+1)'$ and obtain    $P\cdot V= \frac{-1}{12}$, and   massless fields arise without oscillators,
  \dis{
 \begin{array}{ccccc} 
 {s}  & N_i,  & \tilde{s}  \cdot \phi,&\Theta_6, &{\rm Multiplicity}\\[0.3em]
 (\oplus|-+-):&0, &\frac{-1}{12}, & \frac{+6}{12},&2\cdot\sigma_{ 7,8}\\[0.3em] 
 (\ominus|---):&0, & \frac{-5}{12}, & \frac{+10}{12},& 0\cdot\sigma_{7,8} 
  \end{array}  
}
Here,   $Q_{18,20,22}$ charges are
as $(-1,-1,-1)$ for  L-handed fields $\sigma_{5,6}$ and  $(-1,+1,-1)$ for R-handed fields $\sigma_{7,8}$.

\subsection{\bf $\sigma_{11-18}$ from $T_3$ and $T_9$}
In $T_3$, the multiplicity factor is given as
\dis{
\tilde{\chi}(\theta^3,\theta^j)=\left\{\begin{array}{lrrrrrrrrrrrr} 
j=&0,&1,&2,&3,&4,&5,&6,&7,&8,&9,&10,&11 \\
&4,&1,&1,&4,&1,&1,&4,&1,&1,&4,&1,&1 \end{array}\right. 
} 
Since $3\phi=(\frac14,0,\frac14)$, the allowed chiralities are  $s=( \oplus ; \pm,-,\mp)$, and $s=( \ominus;\pm,+,\mp)$.  We have massless conditions for right movers as $\tilde{s}\cdot \phi =0, -\frac13$ for $\oplus$ (R-handed fields),  and   $\tilde{s}\cdot \phi =0, +\frac13$ for $\ominus$ (L-handed fields). 

$P=(0^8)(0^5,0,-1,+1)'$ of $\sigma_{11}$ gives
  $P\cdot V=\frac{+1}{4}$, and we obtain
\dis{
 \begin{array}{ccccc} 
 {s}  & N_i,  & \tilde{s}\cdot \phi,&\Theta_3 ,&{\rm Multiplicity}\\[0.3em]
 (\oplus|+--):&0, & 0, & \frac{+6}{12},& 0\cdot\sigma_{11}\\[0.2em]
 (\oplus|--+):&0, & \frac{-4}{12}, & \frac{+10}{ 12},& 
  0\cdot\sigma_{11} \\[0.2em]
 (\ominus|++-):&0, & \frac{+4}{12}, & \frac{+2}{ 12},& 
 0\cdot\sigma_{11} \\[0.2em]
 (\ominus|-++):&0, & 0, &\frac{+6}{12},& 0\cdot\sigma_{11} 
  \end{array}  
}
To have massless modes, we need additional phases $\frac{\pm 2}{ 12}$ and $\frac{\pm 6}{ 12}$. But,  $\frac{\pm 2}{ 12}$ cannot be used as shown in Table \ref{tab:OsciilatingModes}, and we have the following
\dis{
 \begin{array}{ccccc} 
 {s}  & N_i,  & \tilde{s}\cdot \phi,&\Theta_3,&{\rm Multiplicity}\\[0.3em]
 (\oplus|+--):&\frac{\pm 6}{ 12}, & 0, & \frac{+12,0}{12},& 
 [2(1_{1}+1_{3}),2(1_{\bar{1}}+1_{\bar{3}})]\cdot\sigma_{11,15}\\[0.2em]
 (\oplus|--+):&\frac{\pm 6}{ 12}, & \frac{-4}{12}, & \frac{+4,+4}{ 12},& 
  [1(1_1+1_3),1(1_{\bar{1}}+1_{\bar{3}})]\cdot\sigma_{11,15} \\[0.2em]
 (\ominus|++-):&\frac{\pm 6}{ 12}, & \frac{+4}{12}, & \frac{+8,-4}{ 12},& 
  [1(1_1+1_3),1(1_{\bar{1}}+1_{\bar{3}})]\cdot\sigma_{11,15} \\[0.2em]
 (\ominus|-++):&\frac{\pm 6}{ 12}, & 0, &\frac{+12,0}{12},&   [2(1_{1}+1_{3}),2(1_{\bar{1}}+1_{\bar{3}})]\cdot\sigma_{11,15} 
  \end{array}  
}

$P=(0^5;0,+1,+1)(0^5,0,-1,+1)'$ of $\sigma_{12}$ gives
  $P\cdot V=\frac{-1}{12}$, and we obtain
\dis{
 \begin{array}{ccccc} 
 {s}  & N_i,  & \tilde{s}\cdot \phi,&\Theta_3,&{\rm Multiplicity}\\[0.3em]
 (\oplus|+--):&\frac{\pm 6}{ 12}, & 0, & \frac{+8,-4}{12},& 
 [1(1_{1}+1_{3}),1(1_{\bar{1}}+1_{\bar{3}})]\cdot\sigma_{12,13,16,17}\\[0.2em]
 (\oplus|--+):&\frac{\pm 6}{ 12}, & \frac{-4}{12}, & \frac{+12,0}{ 12},& 
  [1(1_1+1_3),2(1_{\bar{1}}+1_{\bar{3}})]\cdot\sigma_{12,13,16,17} \\[0.2em]
 (\ominus|++-):&\frac{\pm 6}{ 12}, & \frac{+4}{12}, & \frac{+4,-8}{ 12},& 
  [1(1_1+1_3),1(1_{\bar{1}}+1_{\bar{3}})]\cdot\sigma_{12,13,16,17} \\[0.2em]
 (\ominus|-++):&\frac{\pm 6}{ 12}, & 0, &\frac{+8,-4}{12},&   [1(1_{1}+1_{3}),1(1_{\bar{1}}+1_{\bar{3}})]\cdot\sigma_{12,13,16,17} 
  \end{array}  
}

$P=(0^5;+1,+1,0)(0^5,-1,-1,0)'$ of $\sigma_{14}$ gives
  $P\cdot V=\frac{-5}{6}$  and   $\Theta^0=P\cdot V+\frac{k}{12}=\frac{-7}{12}$. So, we have
\dis{
 \begin{array}{ccccc} 
 {s}  & N_i,  & \tilde{s}\cdot \phi,&\Theta_3,&{\rm Multiplicity}\\[0.3em]
 (\oplus|+--):& \frac{-5,-1}{ 12}, & 0, & \frac{-12,-8}{12},& 
 [2(1_{\bar{1}}),1(1_{\bar{3}})]\cdot\sigma_{14,18}\\[0.2em]
 (\oplus|--+):&\frac{-1}{ 12}, & \frac{-4}{12}, & \frac{-12}{ 12},& 
2(1_{\bar{3}}) \cdot\sigma_{14,18} \\[0.2em]
 (\ominus|++-):&\times, & \frac{+4}{12}, & \times,& 
0\cdot\sigma_{14,18} \\[0.2em]
 (\ominus|-++):& \frac{-5,-1}{ 12}, & 0, & \frac{-12,-8}{12},& 
 [2(1_{\bar{1}}),1(1_{\bar{3}})]\cdot\sigma_{14,18}
  \end{array}  
}

We list only L-handed fields in Table II.

\begin{table}[t!]
\begin{center}
\begin{tabular}{@{}c|ccccccc @{}} \botrule
 $2\tilde{c}\,(k=)$&$1$ &$2$  &$3$&$4$ &$5$ &$6$  \\[0.2em] \colrule
 $\Z_{12-I}$  & $-\frac{35}{24}$&$-\frac{3}{2}$& $-\frac{13}{8}$&$-\frac{4}{3}$& $-\frac{35}{24}$&$-\frac{3}{2}$  \\[0.2em]
$\Z_{12-II}$  & $-\frac{103}{72}$&$-\frac{31}{18}$  & $-\frac{11}{8}$ & $-\frac{14}{9}$ & $-\frac{103}{72}$ & $-\frac{3}{2}$   \\[0.2em]
$\Z_{8-I}$  & $-\frac{47}{32}$&$-\frac{11}{8}$ &$-\frac{47}{32}$ & $-\frac{3}{2}$  
\\[0.2em]
$\Z_{8-II}$  & $-\frac{45}{32}$&$-\frac{13}{8}$  & $-\frac{45}{32}$ & $-\frac{3}{2}$\\[0.2em]
$\Z_{7}$  & $-\frac{10}{7}$& $-\frac{10}{7}$ & $-\frac{10}{7}$\\[0.2em]
$\Z_{6-I}$  & $-\frac{3}{2}$&$-\frac{4}{3}$ & $-\frac{3}{2}$  \\  [0.2em]
 $\Z_{6-II}$  & $-\frac{25}{18}$&$-\frac{28}{18}$ & $-\frac{3}{2}$ & & &  \\  [0.2em]
 $\Z_{4}$  & $-\frac{11}{8}$ & $-\frac{3}{2}$  \\   [0.2em]
$\Z_{3}$  & $-\frac{4}{3}$  \\   [0.2em]
 \colrule
$\Z_{2}(6D)$  & $-\frac{3}{2}$  \\   [0.2em]
 \botrule
 \end{tabular} \label{tab:VacE} 
\end{center}
\caption{Two times right-mover vacuum energy $2\tilde{c}$ of Ref. \cite{LNP696}. Typos in  $\Z_{2}(6D)$ of Ref. \cite{LNP696} are corrected here.}
\end{table}

\begin{table}[t!]
\begin{center}
\begin{tabular}{@{}c|ccccccc @{}} \botrule
 $2c\,(k=)$&$1$ &$2$  &$3$&$4$ &$5$ &$6$  \\[0.2em] \colrule
 $\Z_{12-I}$  & $-\frac{11}{24}$&$-\frac{1}{2}$& $-\frac{5}{8}$&$-\frac{1}{3}$& $-\frac{11}{24}$&$-\frac{1}{2}$  \\[0.2em]
$\Z_{12-II}$  & $-\frac{31}{72}$&$-\frac{13}{18}$  & $-\frac{3}{8}$ & $-\frac{5}{9}$ & $-\frac{31}{72}$ & $-\frac{1}{2}$   \\[0.2em]
$\Z_{8-I}$  & $-\frac{15}{32}$&$-\frac{3}{8}$ &$-\frac{15}{32}$ & $-\frac{1}{2}$  
\\[0.2em]
$\Z_{8-II}$  & $-\frac{13}{32}$&$-\frac{5}{8}$  & $-\frac{13}{32}$ & $-\frac{1}{2}$\\[0.2em]
$\Z_{7}$  & $-\frac{3}{7}$& $-\frac{3}{7}$ & $-\frac{3}{7}$\\[0.2em]
$\Z_{6-I}$  & $-\frac{1}{2}$&$-\frac{1}{3}$ & $-\frac{1}{2}$  \\  [0.2em]
 $\Z_{6-II}$  & $-\frac{7}{8}$&$-\frac{5}{9}$ & $-\frac{1}{2}$\\  [0.2em]
 $\Z_{4}$  & $-\frac{3}{8}$ & $-\frac{1}{2}$  \\   [0.2em]
$\Z_{3}$  & $-\frac{1}{3}$  \\   [0.2em]
 \colrule
$\Z_{2}(6D)$  & $-\frac{1}{2}$  \\   [0.2em]
 \botrule
 \end{tabular} \label{tab:VacE} 
\end{center}
\caption{Two times right-mover vacuum energy $2c$ of Ref. \cite{LNP696}. Typos in  $\Z_{6-II}$ of Ref. \cite{LNP696} are corrected here.}
\end{table}

\subsection{\bf $\sigma_{9-10}$ from $T_2^0$}
We have
\dis{
\tilde{\chi}(\theta^2,\theta^j)=\left\{\begin{array}{lrrrrrrrrrrrr} 
j=&0,&1,&2,&3,&4,&5,&6,&7,&8,&9,&10,&11 \\
&3,&3,&3,&3,&3,&3,&3,&3,&3,&3,&3,&3 \end{array}\right..
}
For $T_2$, the masslessness condition of R-sector is $(s+ (2\phi))^2=\frac12$, which is satisfied by $s=(\ominus;---)$. With $P=(0^8)(0^5;-1,0,+1)'$, we have $P\cdot V=\frac14$, and $ \Theta_2^0=\frac{-2}{12}$. We cannot make up $\frac{-3}{12}$ with the modes allowed in Table \ref{tab:OsciilatingModes}.

\subsection{\bf $\sigma_{19-21}$ from $T_1^0$}
For $T_1$, we have the following multiplicity factor 
\dis{
\tilde{\chi}(\theta^1,\theta^j)=\left\{\begin{array}{lrrrrrrrrrrrr} 
j=&0,&1,&2,&3,&4,&5,&6,&7,&8,&9,&10,&11 \\
&3,&3,&3,&3,&3,&3,&3,&3,&3,&3,&3,&3 \end{array}\right..
}
For $T_1^0$, $(s+\phi)^2=\frac{11}{24}$ is satisfied by $s=(\ominus|---)$. With $P=(0^8)(0^5;-1,0,+1)'$ and $P\cdot V=\frac{1}{4}$, Eq. (\ref{Theta0}) gives
   \dis{
 \begin{array}{ccccc} 
 {s}  & N_i,  & \tilde{s}\cdot  \phi ,&\Theta_1^0,&{\rm Multiplicity}\\[0.3em]
 (\ominus|---):&0, & \frac{-5}{12}, & \frac{+9}{ 12},& 0\cdot\sigma_{19,20,21}
  \end{array}  
}
and there is no massless field without oscillators.   We cannot make up $\frac{+3}{12}$ with the modes allowed in Table \ref{tab:OsciilatingModes}.
 
\subsection{\bf $\sigma_{22-24}$ from $T_5^0$}
For $T_5$, we have the following multiplicity factor 
\dis{
\tilde{\chi}(\theta^7,\theta^j)=\left\{\begin{array}{lrrrrrrrrrrrr} 
j=&0,&1,&2,&3,&4,&5,&6,&7,&8,&9,&10,&11 \\
&3,&3,&3,&3,&3,&3,&3,&3,&3,&3,&3,&3 \end{array}\right..
}
For $T_5^0$, $(s+5\phi)^2=\frac{11}{24}$ is satisfied by $s=(\oplus|+--)$ and $(\ominus|-++)$. With $P=(0^5;0,+1,+1)(0^5; -1,-1,0)'$, we have $P\cdot V=\frac{+2}{12}$, Eq. (\ref{Theta0}) gives
   \dis{
 \begin{array}{ccccc} 
 {s}  & N_i,  & \tilde{s}\cdot( \phi),&\Theta_5^0,&{\rm Multiplicity}\\[0.3em]
 (\oplus|+--):&0, &0, & \frac{+4}{ 12},& 0\cdot\sigma_{22,23,24}\\[0.2em]
 (\ominus|-++):&0, &0, & \frac{+4}{ 12},& 0\cdot\sigma_{22,23,24} 
  \end{array}  
} 
and there is no massless field without oscillators.   With the modes allowed in Table \ref{tab:OsciilatingModes},  we obtain
   \dis{
 \begin{array}{ccccc} 
 {s}  & N_i,  & \tilde{s}\cdot( \phi),&\Theta_5^0,&{\rm Multiplicity}\\[0.3em]
 (\oplus|+--):&\frac{-4}{ 12}(1_{\bar{1}}+1_3), &0, & 0,& 1\cdot\sigma_{22,23,24}\\[0.2em]
 (\ominus|-++):&\frac{-4}{ 12}(1_{\bar{1}}+1_3), &0, & 0,& 1\cdot\sigma_{22,23,24} 
  \end{array}  
} 
and  the charges of L-handed fields are 
\dis{
Q_{18,20,22}=(-1,+1,+1).
}

\end{appendix}

\acknowledgments{I thank Junu Jeong and Se-Jin Kim for numerical-checking of the diagonalization processes. This work is supported in part by the National Research Foundation (NRF) grant  NRF-2018R1A2A3074631 and by the IBS (IBS-R017-D1-2014-a00). }

 
\end{document}